\documentclass[10pt]{article}
\usepackage[T1]{fontenc}
\usepackage[latin9]{inputenc}
\usepackage{geometry}
\usepackage{prettyref}
\usepackage{amsmath}
\usepackage{graphicx}
\usepackage{setspace}
\usepackage{amssymb}
\usepackage{esint}

\makeatletter

\usepackage[T1]{fontenc}
\usepackage{ae}
\usepackage{aecompl}

\usepackage[sf,bf,small]{titlesec}

\usepackage[hang,small,bf,sf]{caption} 

\renewcommand{\bar}[1]{\overline{#1}}

\DeclareMathOperator{\re}{Re}  
\DeclareMathOperator{\im}{Im}

\DeclareMathOperator{\sech}{sech}
\DeclareMathOperator{\sign}{sign}

\AtBeginDocument{
  
}

\makeatother

\begin{document}
\begin{flushright}
BROWN-HET 1497\\
arXiv:0803.4222
\par\end{flushright}

~

\begin{doublespace}
\begin{center}
\textsf{\textbf{\LARGE Vibrating Giant Spikes}}\textsf{\textbf{\Large }}\\
\textsf{\textbf{\Large and the large-winding sector}}
\par\end{center}{\Large \par}
\end{doublespace}

~

\begin{onehalfspace}
\begin{center}
Michael C. Abbott and In\^{e}s V. Aniceto\\
\emph{Brown University, Providence RI, USA.}\\
\emph{abbott, nes@het.brown.edu}\\
~\\
 28 March 2008\\
(updated 4 May, v2)
\par\end{center}
\end{onehalfspace}

~

\begin{quote}
\textsf{\textbf{Abstract}}:~ The single spike is a rigidly rotating
classical string configuration closely related to the giant magnon.
We calculate bosonic and fermionic modes of this solution, from which
we see that it is not supersymmetric. It can be viewed as an excitation
above a hoop of string wound around the equator, in the same sense
that the magnon is an excitation above an orbiting point particle.
We find the operator which plays the role of the Hamiltonian for this
sector, which compared to the magnon's $\Delta-J$ has the angular
momentum replaced by a winding charge. The single spike solution is
unstable, and we use the modes to attempt a semi-classical computation
of its lifetime. 
\end{quote}
~

\section{Introduction}

Much has been learned about the AdS-CFT correspondence, which relates
large-$N$ gauge theory to string theory \cite{Maldacena:1997re},
by looking at limits in which an $SO(6)$ charge $J$ also becomes
large. At large $\lambda$ the theory is a theory of classical strings
moving in $AdS_{5}\times S^{5}$, with $J$ an angular momentum on
the sphere, while at small $\lambda$ it is perturbative Yang--Mills
theory in 4 dimensions, with $J$ an R-charge of this theory. \cite{Gubser:2002tv,Minahan:2006bd}
This is the large-$J$ sector of the correspondence.

The first well-studied example in this sector is the BMN limit \cite{Berenstein:2002jq,Plefka:2003nb}
which on the string side, consists of nearly point-like solutions
orbiting the sphere, experiencing a pp-wave geometry. On the gauge
theory side, the anomalous dimension $\Delta-J$ can be computed as
the energy of a ferromagnetic spin chain. \cite{Hofman:2006xt,Beisert:2003ys,Beisert:2005tm}
These spin chains are integrable systems, allowing the use of Bethe
ansatz techniques to compute the spectrum from the S-matrix for two-particle
scattering. \cite{Minahan:2002ve,Beisert:2003xu,Beisert:2003ea,Kruczenski:2004kw,Staudacher:2004tk,Beisert:2006qh}
(In some cases one can explicitly recover the string action from the
spin-chain \cite{Kruczenski:2003gt}.) 

The elementary excitations of spin chains are magnons, which to be
scattered must have some momentum $p\neq0$. Extending the theory
to allow lone magnons with momentum leads to the centrally extended
algebras \cite{Beisert:2005tm,Arutyunov:2006ak} on the gauge side,
dual to strings which do not close. These are the giant magnons \cite{Hofman:2006xt}.
Generalisations which have been explored include magnons with more
than one large angular momentum \cite{Dorey:2006dq,Chen:2006gea,Ryang:2006yq}
and magnons with finite $J$ \cite{Arutyunov:2006gs,Okamura:2006zv,Astolfi:2007uz,Klose:2008rx}.

Giant magnons are one type of rigidly rotating strings with cusps,
moving on the sphere and made as large as they can be. In general
these are called spiky strings, and they also exist in flat space
\cite{Burden:1984xk,Burden:1985md} and in $AdS$. \cite{Kruczenski:2004wg,Jevicki:2007aa}
In flat space T-duality leads to another class of spiky strings, with
cusps pointing inwards, and these `T-dual' solutions can also exist
on the sphere. Starting with one of these and taking the same maximum-size
limit used for the magnon then leads to the single spike solution
which we study here. \cite{Ishizeki:2007we} Recent papers on the
single spike include \cite{Kruczenski:2006pk,Mosaffa:2007ty,Ishizeki:2007kh,Bobev:2007bm,Bozhilov:2007wn,Dimov:2007ey,Kluson:2007fr}. 

The giant magnon can be viewed as an excitation above a vacuum solution
of a point particle orbiting along the equator. \cite{Spradlin:2006wk}
(The label `giant' is meant to indicate that they explore much of
the $S^{5}$ geometry, as the earlier giant gravitons did. \cite{McGreevy:2000cw,Corley:2001zk})
Fluctuations of this vacuum have Hamiltonian $\Delta-J$ \cite{Frolov:2002av}
(where $J\leq\Delta$ is the BPS bound.) The single spike is similarly
an excitation above a string wound around the equator, which we call
the hoop. In the Hamiltonian for fluctuations, the angular momentum
$J$ is replaced with a measure of the winding along the same direction,
which we call $\Phi$. This is almost T-duality, except that the circle
involved is part of sphere. It is not clear whether this duality can
be usefully related to the T-duality used in \cite{Lunin:2005jy}
and \cite{Alday:2007hr}, in $S^{5}$ and $AdS$.

The single spike, and indeed the hoop, are not supersymmetric. Exploring
the correspondence in sectors with less or no supersymmetry is of
great interest, and it is our hope that the close relationship to
the magnon case can be used as a tool for this. The gauge theory dual
of the single spike is not known, but it is conjectured to be some
excitation of an anti-ferromagnetic state of the spin chain \cite{Roiban:2006jt,desCloizeaux62}
in what has been named the large-winding sector of the correspondence
\cite{Hayashi:2007bq}. In the absence of supersymmetry it is possible
that integrability will help to find the dual of the spike solution. 

Solitons have long been studied in field theory, and a set of tools
called semi-classical quantisation enables us to learn about the related
objects in the quantum theory. \cite{Dashen:1975hd,deVega:1982sh,Gervais:1976wr,Faddeev:1977rm,Rajaraman:1975ez,Scott:1973eg,Coleman:1988}
Many of these techniques have been revived to study solutions of classical
string theory in $AdS_{5}\times S^{5}$ \cite{Frolov:2002av,Frolov:2003tu,Papathanasiou:2007gd,Chen:2007vs}
(which is known to be integrable \cite{Bena:2003wd,Mandal:2002fs}).
We have the extra complication that the single spike is an excitation
of an unstable vacuum state (as the string wrapped around an equator
of $S^{5}$ can slide off towards the pole) so what we aim to calculate
by these methods is not an energy correction but a lifetime, as discussed
in the text.

\subsubsection{Contents}

In section \ref{sec:Bosonic-Sector} we set out the solution, and
write down its bosonic modes. The bosonic string on the sphere can
be mapped to sine-gordon theory, where the giant magnon becomes the
simple kink. The single spike is instead mapped to an unstable kink. 

Section \ref{sec:Fermionic-Sector} contains the calculation of the
ferminonic modes, along the lines of what was done for the giant magnon
in \cite{Minahan:2007gf} and \cite{Papathanasiou:2007gd}. We find
that, compared to the magnon, some of the bosonic modes become tachyonic,
while the fermionic modes double in number and become massless, ruling
out supersymmetry.

In section \ref{sec:Section-Four} we study the vacuum of the large-winding
sector, the infinitely wound hoop. We find that $\Delta-\Phi$ is
the Hamiltonian for perturbations of this hoop, and therefore is the
charge which receives quantum corrections. After making a rough calculation
of these corrections we conclude in section \ref{sec:Conclusion}.

Appendix \ref{sec:Fermionic-Zero-modes} is a computation of fermionic
zero modes, which are in this case the $\omega\to0$ end of the continuum
of non-zero modes. Appendix \ref{sec:Appendix-Energy-Corr} has some
further details on quantum corrections.

\section{Bosonic Sector\label{sec:Bosonic-Sector}}

\subsection{Spiky strings in flat space\label{sub:flat-space}}

\begin{figure}
\begin{centering}
~ ~ \includegraphics[width=0.3\columnwidth]{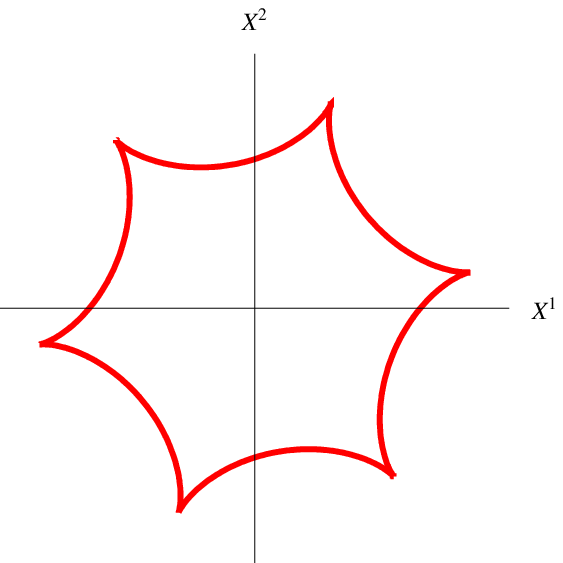} ~ ~ \includegraphics[width=0.3\columnwidth]{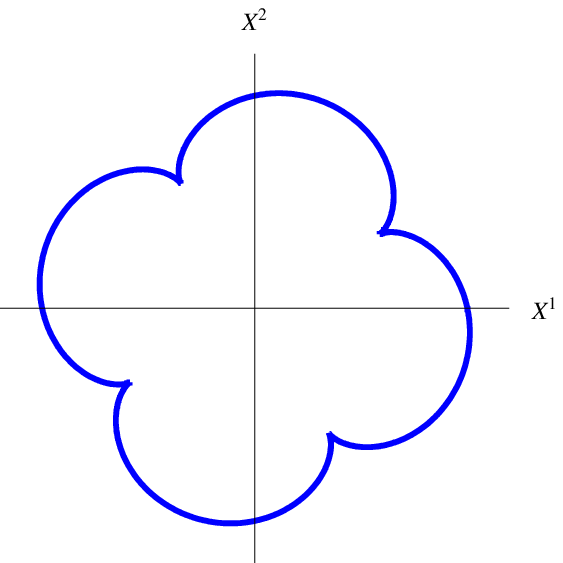}
\par\end{centering}

\caption{The original and T-dual spiky string in flat space. Both are drawn
with $B=5$, leading to 6 and 4 spikes respectively. \label{fig:Flat-space}}

\end{figure}

The spiky string in flat space is the solution \cite{Kruczenski:2004wg}\cite{Burden:1984xk,Burden:1985md}\begin{align}
X^{0} & =t,\nonumber \\
X^{1} & =A\cos\left(\frac{t+x}{2A}\right)+AB\cos\left(\frac{t-x}{2AB}\right),\label{eq:flat-original}\\
X^{2} & =A\sin\left(\frac{t+x}{2A}\right)+AB\sin\left(\frac{t-x}{2AB}\right),\nonumber \end{align}
with two parameters $A,B$. This describes a rigidly rotating string
with $n=B+1$ cusps, or spikes, pointing outwards (see figure \ref{fig:Flat-space}).
$A$ clearly controls the overall size. 

Since this solution has neither centre-of-mass momentum nor winding,
the effect of T-duality in the $X^{2}$ direction is to change the
sign of the left-movers in that direction, \cite{Ishizeki:2007we}
giving \begin{align}
 & X^{0}\mbox{ and }X^{1}\mbox{ unchanged,}\label{eq:flat-t-dual}\\
X^{2} & =A\sin\left(\frac{t+x}{2A}\right)-AB\sin\left(\frac{t-x}{2AB}\right),\nonumber \end{align}
which is another rigidly rotating string, now with $B-1$ spikes pointing
inwards. In both cases the cusps always move at the speed of light.

Notice that the T-dual solution could be obtained by simply interchanging
$x$ and $t$ in the spatial $X^{i}$. This symmetry is visible in
the equations of motion\[
\left(-\partial_{t}^{2}+\partial_{x}^{2}\right)X^{i}=0\]
and in the Virasoro constraints (for $X^{0}=t$) \[
\left(\partial_{t}X^{i}\right)^{2}+\left(\partial_{x}X^{i}\right)^{2}=1,\qquad\partial_{t}X^{i}\partial_{x}X^{i}=0,\]
all of which are unchanged by $x\leftrightarrow t$.

\subsection{On the sphere}

Similar solutions exist on the sphere, and when they are small will
reduce to those in flat space. In \cite{Ishizeki:2007we} it was shown
that if the analogue of the original solution \eqref{eq:flat-original}
becomes large, so that the spikes touch the equator, then each segment
(between spikes) of it becomes a giant magnon. For the analogue of
the T-dual solution \eqref{eq:flat-t-dual}, the limit in which the
lobes touch the equator is the single spike which this paper studies.

We embed the sphere in $\mathbb{R}^{6}$, parameterised by $X^{i}$
(with $X^{i}X^{i}=1$), and look at solutions rotating in the $Z_{1}=X^{1}+iX^{2}$
plane. The remaining four directions we call $\vec{X}$, and $X^{0}$
is the time co-ordinate (ultimately from $AdS$).

The giant magnon \cite{Hofman:2006xt} is the following solution:
\begin{align}
X^{0} & =t,\nonumber \\
Z_{1} & =e^{it}\left(c+i\sqrt{1-c^{2}}\tanh u\right),\label{eq:TheMagnon}\\
\vec{X} & =\vec{n}\sqrt{1-c^{2}}\sech u,\nonumber \end{align}
where we write $c=\cos(p/2)$ for the worldsheet velocity, and $(u,v)$
are boosted worldsheet co-ordinates\begin{align}
u & =\gamma(x-ct),\label{eq:defn-uv}\\
v & =\gamma(t-cx),\qquad\mbox{with }\gamma=\frac{1}{\sqrt{1-c^{2}}}=\frac{1}{\sin(p/2)}.\nonumber \end{align}
Note all of $-\infty<x<\infty$ covers only one of the curves between
cusps. It is understood that the physical closed-string solution consists
of several giant magnons connected together. The case $c=0$ (zero
worldsheet velocity, $p=\pi)$ is one of GKP's folded strings. \cite{Gubser:2002tv}
In the limit $p\to0$ the magnon becomes a point particle moving along
the equator.

\begin{figure}
\begin{centering}
\includegraphics[width=0.3\columnwidth]{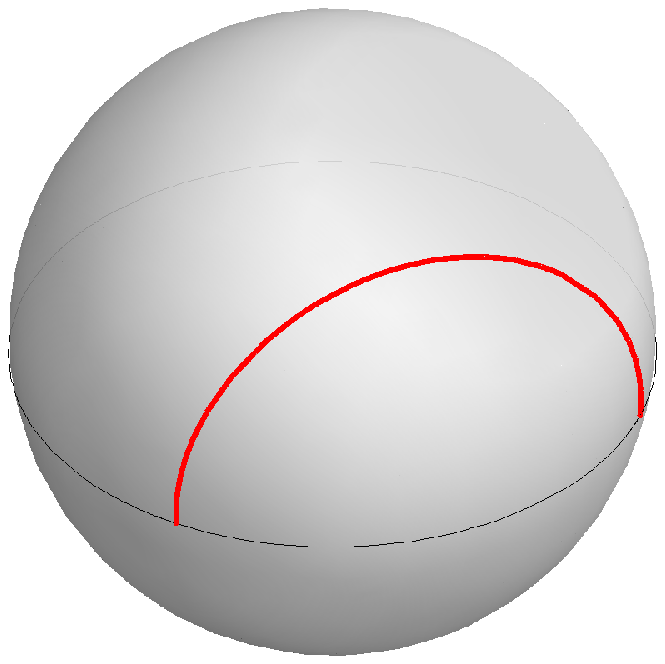} ~
~ \includegraphics[width=0.3\columnwidth]{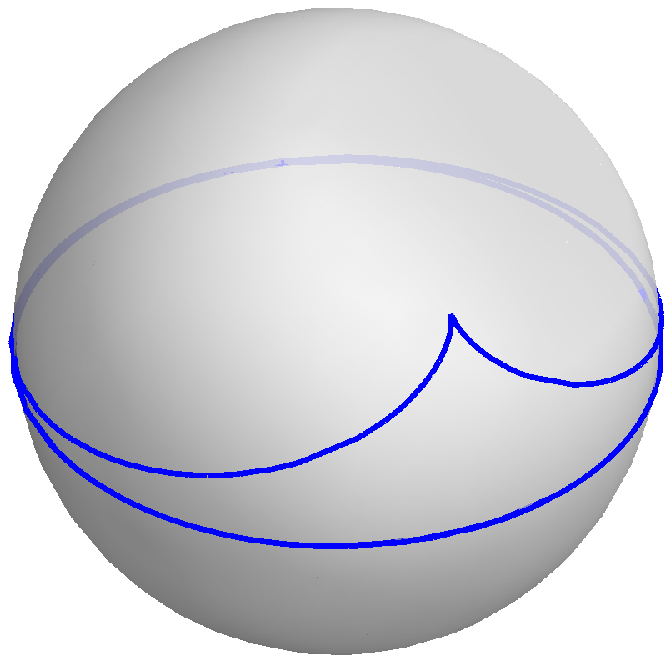}
\par\end{centering}

\caption{The giant magnon (left, $c=\cos(p/2)=0.7$) and the single spike (right,
$c=0.8$). These are both are rigidly rotating along the equator,
with their cusps moving at the speed of light. \label{fig:Sphere-plots}}

\end{figure}

This solution is written in conformal gauge (i.e. the induced metric
is proportional to the standard metric, $\partial_{a}X^{\mu}\partial_{b}X^{\nu}\eta_{\mu\nu}\propto\eta_{ab}$)
and thus solves the Virasoro constraints \[
\left(\partial_{t}X^{i}\right)^{2}+\left(\partial_{x}X^{i}\right)^{2}=1,\qquad\partial_{t}X^{i}\partial_{x}X^{i}=0,\]
and the conformal gauge equations of motion\[
\left(-\partial_{t}^{2}+\partial_{x}^{2}\right)X^{i}+X^{i}\left(-(\partial_{t}X^{j})^{2}+(\partial_{x}X^{j})^{2}\right)=0.\]
(The extra term, compared to flat space, comes from the constraint
$X^{i}X^{i}=1$.) As in flat space, these are unchanged by the interchange
of $x$ and $t.$ So there is another solution $X_{\mathrm{spike}}^{i}(t,x)=X_{\mathrm{magnon}}^{i}(x,t)$,
which has been dubbed the single spike: \cite{Ishizeki:2007we} \begin{align}
X^{0} & =t,\nonumber \\
Z_{1} & =e^{ix}\left(c+i\sqrt{1-c^{2}}\tanh v\right),\label{eq:TheSpike}\\
\vec{X} & =\vec{n}\sqrt{1-c^{2}}\sech v\,.\nonumber \end{align}
This solution is drawn in figure \ref{fig:Sphere-plots}. We keep
the same parameter $c$, although the worldsheet velocity is now $1/c$
in the $x,t$ co-ordinates.%
\footnote{This is related to the parameter $\theta_{0}$ used in \cite{Ishizeki:2007we},
which is the angle from the north pole to the tip of the spike, by
$\sin\theta_{0}=c=\cos(p/2)$. Also note that $\bar{\theta}=\frac{\pi}{2}-\theta_{0}=p/2$.%
} For the most part we will be interested only in the range $0<c<1$.

Both solutions are localised on the worldsheet. As $x\to\infty$,
the magnon solution approaches the point particle $Z_{1}=e^{it}$
and $\vec{X}=0$ while the single spike solution becomes instead the
infinitely wound hoop $Z_{1}=e^{ix}$. The point particle and the
hoop are clearly related by the same $x\leftrightarrow t$ swop, and
they are also the vacuum solutions needed to obtain the magnon or
the single spike by the dressing method, which survives this interchange.
\cite{Zakharov:1973pp,Mikhailov:2005zd}\cite{Spradlin:2006wk,Ishizeki:2007kh}

The string's conserved charges of interest are defined as\begin{align}
\Delta & =\frac{\sqrt{\lambda}}{2\pi}\int dx\:1 &  & \mbox{energy (simple in this gauge!),}\nonumber \\
J & =\frac{\sqrt{\lambda}}{2\pi}\int dx\:\im\left(\overline{Z}_{1}\partial_{t}Z_{1}\right) &  & \mbox{angular momentum in }Z_{1}\mbox{ plane,}\nonumber \\
\Phi & =\frac{\sqrt{\lambda}}{2\pi}\int dx\:\im\left(\partial_{x}\log Z_{1}\right)=\frac{\sqrt{\lambda}}{2\pi}\Delta\phi &  & \mbox{winding charge.}\label{eq:winding-defn}\end{align}
This $\Phi$ is a conveniently scaled version of the the opening angle
$\Delta\phi$, where $\phi=\arg Z_{1}$ is the azimuthal angle. 

For the magnon, $\Delta$ and $J$ are infinite, and we have the familiar
\begin{align}
\Delta-J & =\frac{\sqrt{\lambda}}{\pi}\sin(p/2),\label{eq:charges-for-magnon}\\
\Phi & =\frac{\sqrt{\lambda}}{2\pi}p\,.\nonumber \end{align}
For the single spike, it is $\Phi$ instead of $J$ which is infinite,
and \begin{align}
\Delta-\Phi & =\frac{\sqrt{\lambda}}{2\pi}p,\label{eq:charges-for-spike}\\
J & =\frac{\sqrt{\lambda}}{\pi}\sin(p/2)\,.\nonumber \end{align}

\subsection{Zero modes\label{sub:bosonic-zero-modes}}

Bosonic zero modes are the variations produced by changing collective
co-ordinates: \[
\delta_{v}X^{i}=-\frac{\partial X^{i}}{\partial v_{0}}\Big\vert_{v_{0}=0},\]
where $v_{0}$ is some modulus. Writing the single spike solution
\eqref{eq:TheSpike} with explicit $x_{0}$ and $v_{0}$ in addition
to the direction $\vec{n}$ \begin{align*}
Z_{1} & =e^{i(x-x_{0})}\left(c+i\sqrt{1-c^{2}}\tanh(v-v_{0})\right),\\
\vec{X} & =\vec{n}\sqrt{1-c^{2}}\sech(v-v_{0}),\end{align*}
we obtain the following modes: 

\begin{itemize}
\item $\delta_{x}$, a rigid rotation of $Z_{1}$:\begin{align*}
\delta_{x}Z_{1} & =iZ_{1}\,,\\
\delta_{x}\vec{X} & =0\,;\end{align*}

\item $\delta_{v}$, a reparametrisation along $v$:\begin{align}
\delta_{v}Z_{1} & =e^{it}i\sqrt{1-c^{2}}\sech^{2}v\,,\label{eq:delta-v-zero-mode}\\
\delta_{v}\vec{X} & =-\vec{n}\sqrt{1-c^{2}}\sech v\,\tanh v\,;\nonumber \end{align}

\item $\delta_{m}$, a rotation of the orientation $\vec{n}$: \begin{align*}
\delta_{m}Z_{1} & =0\,,\\
\delta_{m}\vec{X} & =\vec{m}\,\sqrt{1-c^{2}}\sech v\,,\end{align*}
where $\vec{m}\cdot\vec{n}=0$, thus there are three such modes. 
\end{itemize}
It may seem strange to work out $\delta_{v}$ holding $x$ fixed (and
$\delta_{x}$ holding $v$ fixed) rather than always using one pair
$x,t$ or $u,v$. This simply gives a convenient linear combination
of the modes, in which one is normalisable and one is not. If instead
we worked out $\delta_{t}$ holding $x$ fixed (and \emph{vice versa})
we would get\begin{align*}
\delta_{t\vert x}X^{i} & =\gamma\delta_{v}X^{i} & \delta_{x\vert t}X^{i} & =\delta_{x}X^{i}-c\gamma\delta_{v}X^{i}\\
\delta_{t\vert x}X^{0} & =1 & \delta_{x\vert t}X^{0} & =0\end{align*}
where we now write the time components in addition to the spatial
ones. In spacetime the meaning of these two modes ($\delta_{t\vert x}$
and $\delta_{x\vert t}$) is clear: at any point they are the two
tangent vectors to the string. In fact they are exactly the co-ordinate
basis vectors from $x,t$. These would normally generate reparametrisations,
not physical modes. 

But here, as for the giant magnon, we are not studying the complete
string solution, but rather just a section of it.%
\footnote{We return to this question in section \ref{sub:About-these-boundary}
below.%
} There must be solitons elsewhere on the worldsheet, and motion relative
to these is physical. Thus we keep one of these modes, along with
the 3 modes $\delta_{m}$, making 4 zero modes in total.%
\footnote{Our mode $\delta_{v}$ \eqref{eq:delta-v-zero-mode} is the analogue
of \cite{Minahan:2007gf}'s (3.11) and \cite{Papathanasiou:2007gd}'s
(2.16). In \cite{Minahan:2007gf} this is derived from a translation
of the sine-gordon soliton.%
} 

Notice that the other physical zero modes, the perpendicular rotations
$\delta_{m}$, are independent of $u$. The corresponding modes in
the magnon case, (2.15) in \cite{Papathanasiou:2007gd}, are independent
of $v$, which is time boosted by $c$. This is a reason for regarding
$u$ as being the time co-ordinate for the purpose of identifying
zero and non-zero modes.%
\footnote{We could regard $u$ as being the product of a boost by velocity $\frac{1}{c}>1$:\begin{equation}
u=\gamma(x-ct)=-\frac{t-\frac{1}{c}x}{\sqrt{(\frac{1}{c})^{2}-1}}\,.\label{eq:u-boost-1overc}\end{equation}
See also equation \eqref{eq:E-s.g-spike} below.%
}

\subsection{Non-zero modes\label{sub:bosonic-nonzero-modes}}

Inserting $X^{i}+\delta X^{i}$ into the equation of motion, we obtain
the equation for the fluctuations\[
\partial_{a}\partial^{a}\delta X^{i}+\left(1-2\sech^{2}v\right)\delta X^{i}-\left(X^{j}\partial_{a}\partial^{a}\delta X^{j}\right)X^{i}=0.\]
The zero modes above are solutions of this equation. We now seek non-zero
modes, i.e. solutions of the form\[
\delta X^{j}=e^{ikv-i\omega u}\, f^{j}(v).\]
The equivalent problem for the magnon was solved in \cite{Papathanasiou:2007gd},
by a method involving finding a scattering solution and analytically
continuing it. Like the background solution, the modes can be read
off by interchanging $x$ and $t$. 

\begin{itemize}
\item First, there is one massless solution (i.e. $\omega^{2}=k^{2}$):\begin{align}
\delta_{r}\vec{X} & =e^{ikv-i\left|k\right|u}\,\vec{n}\left(k+\left|k\right|\cos\frac{p}{2}\right)\sech v\tanh v\,,\label{eq:delta-r}\\
\delta_{r}X^{1}+i\delta_{r}X^{2} & =-i\, e^{ikv-i\left|k\right|u}e^{ix}\left(k-\left|k\right|\sinh v\sinh(v+i\frac{p}{2})\right)\sech^{2}v\,,\nonumber \\
\delta_{r}X^{1}-i\delta_{r}X^{2} & =i\, e^{ikv-i\left|k\right|u}e^{-ix}\left(k-\left|k\right|\sinh v\sinh(v-i\frac{p}{2})\right)\sech^{2}v\,.\nonumber \end{align}
This solution we drop on the grounds that it is pure gauge: at any
given point $(x,t)$, it is just a linear combination of the zero
modes $\delta_{v}$ and $\delta_{x}$, which we showed to be just
reparametrisations.%
\footnote{Note that the breaking of translational symmetry on the worldsheet
(discussed in section \ref{sub:bosonic-zero-modes}) affects only
the zero modes.%
} The required combination is \[
\delta_{r}X^{i}=e^{ikv-i\left|k\right|u}\left(-(k+k\cos\frac{p}{2})\delta_{v}X^{i}+\left|k\right|\delta_{x}X^{i}\right).\]

\item Second, there are three orthogonal fluctuations, in directions $\vec{m}$
with $\vec{m}\cdot\vec{n}=0$:\begin{align}
\delta_{\bot}\vec{X} & =e^{ikv-i\omega u}\,\vec{m}\left(k+i\tanh v\right)\,,\label{eq:delta-perp}\\
\delta_{\bot}X^{1} & =\delta X_{\perp}^{2}=0,\nonumber \end{align}
and one parallel fluctuation, along the spike's orientation $\vec{n}$:\begin{align}
\delta_{||}\vec{X} & =e^{ikv-i\omega u}\,\vec{n}\left(k+i\tanh v-\left(k+\omega\cos\frac{p}{2}\right)\sech^{2}v\right)\,,\label{eq:delta-par}\\
\delta_{||}X^{1}+i\delta_{||}X^{2} & =-i\, e^{ikv-i\omega u}e^{ix}\left(k\sinh v+\omega\sinh(v+i\frac{p}{2})+i\cosh v\right)\sech^{2}v\,,\nonumber \\
\delta_{||}X^{1}-i\delta_{||}X^{2} & =i\, e^{ikv-i\omega u}e^{-ix}\left(k\sinh v+\omega\sinh(v-i\frac{p}{2})+i\cosh v\right)\sech^{2}v\,.\nonumber \end{align}
These all have $\omega^{2}=k^{2}+1$, the dispersion relation for
a massive particle $m^{2}=1$. 
\end{itemize}
Despite appearing massive in $u,v$, these modes nevertheless represent
an instability with respect to physical time. Re-write the modes in
the original co-ordinates $x,t$, by defining the $K,W$ as follows:\begin{equation}
\delta X^{j}=e^{ikv-i\omega u}\, f^{j}(v)=e^{iKx-iWt}\: f^{j}(\gamma(t-cx)).\label{eq:defn-KW}\end{equation}
Then $W^{2}=K^{2}-1$: written in $x,t$, these modes are tachyonic
(with $m^{2}=-1$). In our gauge, $t=X^{0}$ is the target-space's
time coordinate. Since we have no reason to exclude $\left|K\right|<1$,
we have modes with imaginary $W$, which are exponentially growing
or dying in time, rather than oscillating.

\subsection{Modes in AdS directions}

The solutions above are in the $\mathbb{R}\times S^{5}$ subspace
of $AdS_{5}\times S^{5}$. There are no zero modes in the $AdS$ directions,
as the centre is a special place, but there are non-zero modes. These
are simply the modes of a point particle about the centre of Anti-de~Sitter
space, identical to the giant magnon case.

Write the $AdS_{5}$ part of the metric as \[
ds_{\mathrm{AdS}}^{2}=-\left(\frac{1+\eta^{2}/4}{1-\eta^{2}/4}\right)^{2}d\tau^{2}+\frac{1}{\left(1-\eta^{2}/4\right)^{2}}d\eta_{k}d\eta_{k}\,,\]
where $k=1,2,3,4$. In these co-ordinates the modes are simply\[
\eta_{k}(x,t)=e^{iKx-iWt}f_{k}(K)\]
with $W^{2}=K^{2}+1$. (The infinitely wound hoop has identical $AdS$
modes. We write the Lagrangian for perturbations of this in section
\ref{sub:Corrections-to-what?} and all the modes in appendix \ref{sub:hoop-modes}.)

\subsection{The Pohlmeyer map}

The theory of classical strings moving on $\mathbb{R}\times S^{2}$
is equivalent to the sine-gordon model. The mapping goes as follows:
if $X^{\mu}(x,t)$ is a conformal-gauge string solution with $X^{0}=t$,
then the field $\alpha(x,t)$ defined by \cite{Pohlmeyer:1975nb}\begin{align}
\cos\alpha & =-\partial_{t}X^{i}\partial_{t}X^{i}+\partial_{x}X^{i}\partial_{x}X^{i}\label{eq:pohlmeyer-map}\end{align}
obeys the sine-gordon equation\[
-\partial_{t}\partial_{t}\alpha+\partial_{x}\partial_{x}\alpha=\sin\alpha\,.\]
This is the equation of motion for a field with Lagrangian \[
\mathcal{L}=-\frac{1}{2}(\partial_{t}\alpha)^{2}+\frac{1}{2}(\partial_{x}\alpha)^{2}+U(\alpha)\]
(and thus Hamiltonian $\mathcal{H}=\frac{1}{2}(\partial_{t}\alpha)^{2}+\frac{1}{2}(\partial_{x}\alpha)^{2}+U(\alpha)$,
in our sign convention) using potential \[
U(\alpha)=1-\cos\alpha=2\sin^{2}\left(\frac{\alpha}{2}\right)\,.\]

The giant magnon \eqref{eq:TheMagnon} is mapped to the simple kink
\cite{Hofman:2006xt}\begin{align*}
\alpha & =4\arctan\left(e^{-\gamma(x-ct)}\right)\end{align*}
connecting $\alpha=0$ and $\alpha=2\pi$ at $x=\pm\infty$. The point
particle is mapped to the vacuum $\alpha=0$, and the constant in
$U(\alpha)$ was chosen to make its energy (and thus the energy density
away from the kink) zero. Then the kink has energy \begin{equation}
E_{s.g}=8\gamma=\frac{8}{\sin(p/2)}\,.\label{eq:E-s.g-magnon}\end{equation}
The velocity $c$ can be changed by boosting the kink, and the energy
$E_{\mathrm{s.g.}}$ changes as one would expect for a relativistic
object of rest mass 8.%
\footnote{However, giant magnons of different $c$ are not related by worldsheet
boosts (which are just reparametrisations) since $X^{0}=t$ is held
fixed. %
} But this energy, from the sine-gordon model's Hamiltonian, is inverse
to the spin-chain energy constructed out of target space charges $\Delta-J=\frac{\sqrt{\lambda}}{2}\sin(p/2)$.
This mismatch leads to the following difference: while the time-delay
of scattering giant magnons (on the string worldsheet) or kinks (in
sine-gordon theory) is the same, the resulting phase shift of the
wave-functions is different. \cite{Hofman:2006xt} That is because
these two theories are only identical at the classical level. \cite{Mikhailov:2005sy} 

\begin{figure}
\begin{centering}
\includegraphics[width=0.6\columnwidth]{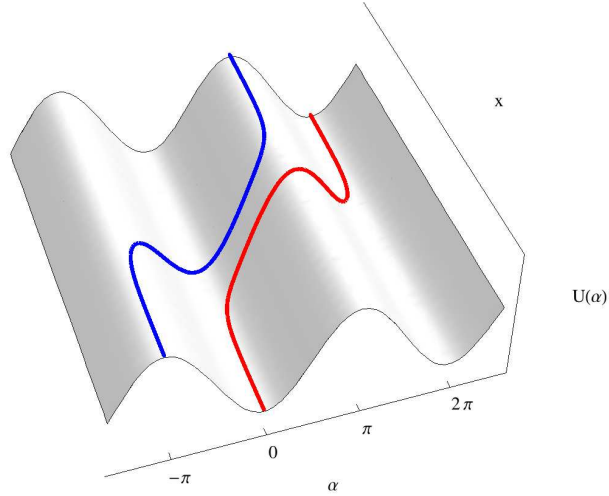} 
\par\end{centering}

\caption{Under the Pohlmeyer map, the magnon is sent to the ordinary kink (in
red) while the single spike is mapped to an unstable solution connecting
the hilltops (in blue). The sine-gordon field $\alpha$ is plotted
left-to-right, $x$ into the page, and $U(\alpha)$ vertically.\label{fig:Pohlmeyer-waves}}

\end{figure}

The single spike \eqref{eq:TheSpike} is mapped instead to an unstable
kink. From the map \eqref{eq:pohlmeyer-map} it is clear that the
effect of the $x\leftrightarrow t$ interchange is to shift the field
by $\pi$:\begin{align*}
\alpha(x,t) & =\alpha_{\mathrm{magnon}}(t,x)-\pi=4\arctan\left(e^{-\gamma(t-cx)}\right)-\pi\,.\end{align*}
This solution connects two adjacent maxima of $U(\alpha)$, rather
than two minima: $\alpha=\pm\pi$ at $x=\pm\infty$. Both cases are
drawn in figure \ref{fig:Pohlmeyer-waves}. If we choose the constant
in $U(\alpha)$ to place these maxima at zero\[
U(\alpha)=-1-\cos\alpha=-2\sin^{2}\left(\frac{\alpha+\pi}{2}\right),\]
then this unstable kink solution has energy\begin{equation}
E_{\mathrm{s.g.}}=8\, c\gamma=\frac{8\cos(p/2)}{\sin(p/2)}=\frac{8}{\sqrt{(\frac{1}{c})^{2}-1}}.\label{eq:E-s.g-spike}\end{equation}

\section{Fermionic Sector\label{sec:Fermionic-Sector}}

To check for supersymmetry, we now calculate the fermionic fluctuations
of this solution. We find that these are all massless, while 2D supersymmetry
would require them to have the same masses as the bosonic modes above.
Also, there are twice as many fermionic as bosonic modes, while supersymmetry
needs equally many.

The calculation follows what was done for the giant magnon by Minahan
\cite{Minahan:2007gf} (zero modes) and Papathanasiou and Spradlin
\cite{Papathanasiou:2007gd} (non-zero modes).

\subsection{Setup }

We follow the notation from \cite{Minahan:2007gf} as much as possible,
except for the worldsheet co-ordinates: we use $(x,t)$ and boost
by $c$ to $(u,v)$ (instead of $(\sigma,t)$ and boost by $v$ to
$(x,\xi)$). Indices $a,b=0,1$ are worldsheet directions, $\mu,\nu$
curved spacetime, $A,B,C$ flat spacetime, and $I,J=1,2$ number fields. 

The unperturbed solution \eqref{eq:TheSpike} lives in $\mathbb{R}\times S^{2}$,
for which we now use co-ordinates $t$ and the usual angles $\theta$
and $\phi$. This part of the metric is then \begin{equation}
g_{\mu\nu}=E_{\mu}^{A}E_{\nu}^{B}\eta_{AB}=\left[\begin{array}{ccc}
-1\\
 & 1\\
 &  & \sin^{2}\theta\end{array}\right]\mbox{ for }\nu=\begin{array}{c}
t\\
\theta\\
\phi\end{array},\label{eq:S2-metric}\end{equation}
so the vielbein's components are $E_{t}^{t}=E_{\theta}^{\theta}=1$
and $E_{\phi}^{\phi}=\sin\theta$. (We are using labels $t,\theta,\phi$
for both curved and flat indices.) The single spike \eqref{eq:TheSpike}
in these co-ordinates is\begin{align*}
X^{0} & =t\,,\\
X^{\theta} & =\theta=\arccos\left(\frac{1}{\gamma\cosh v}\right),\qquad\qquad\mbox{i.e. }\cos\theta=\sqrt{1-c^{2}}\sech v\,,\\
X^{\phi} & =\phi=x+\arctan\left(\frac{\tanh v}{c\gamma}\right),\end{align*}
where $u,v,\gamma$ are still given by \eqref{eq:defn-uv}.

The fermionic fluctuations are two Majorana--Weyl fields $\Theta^{I}$,
with action given by Metsaev and Tseytlin \cite{Metsaev:1998it}%
\footnote{Note that we use $\epsilon$ and $\eta$ with different kinds of indices:
$\epsilon^{ab=01}=1=\epsilon^{AB=12}$, and $\eta^{ab=00}=-1=\eta^{IJ=11}$.
Our gamma-matrices are in the all imaginary basis: $\Gamma_{A\neq0}$
are Hermitian and $\Gamma_{0}$ is anti-Hermitian. $\Gamma_{AB}=\Gamma_{[A}\Gamma_{B]}$,
thus $\Gamma_{\phi\theta}=\Gamma_{\phi}\Gamma_{\theta}$. %
}\[
S=2\frac{\sqrt{\lambda}}{4\pi}\int dtdx\;\mathcal{L}_{F}\qquad\mbox{ where }\quad\mathcal{L}_{F}=i(\eta^{ab}\delta^{IJ}+\epsilon^{ab}\eta^{IJ})\;\overline{\Theta}^{I}\rho_{a}D_{b}\theta^{J}\,.\]
The covariant derivative is defined as \[
D_{a}\Theta^{I}=\left(\partial_{a}+\frac{1}{4}\omega_{a}^{AB}\Gamma_{AB}\right)\delta^{IJ}\Theta^{J}-\frac{i}{2}\Gamma_{\star}\rho_{a}\epsilon^{IJ}\Theta^{J}\]
where $\Gamma_{\star}=i\Gamma_{01234}=i\Gamma_{[0}\Gamma_{1}\Gamma_{2}\Gamma_{3}\Gamma_{4]}$
(these are the $AdS$ directions) has $\Gamma_{\star}^{2}=1$. This
action leads to the following equations of motion: \begin{align*}
(\rho_{0}-\rho_{1})\left(D_{0}+D_{1}\right)\Theta^{1} & =0\,,\\
(\rho_{0}+\rho_{1})\left(D_{0}-D_{1}\right)\Theta^{2} & =0\,.\end{align*}
The projections of the gamma matrices $\rho_{a}=\Gamma_{A}E_{\mu}^{A}\partial_{a}X^{\mu}$
and the spin connection $\omega_{a}^{AB}=\omega_{\mu}^{AB}\partial_{a}X^{\mu}$
are:%
\footnote{As functions of $\theta$, these are simply related to their cousins
in the giant magnon case: $\rho_{0}=\rho_{1}^{\mathrm{magnon}}+\Gamma_{0}$,
$\rho_{1}=\rho_{0}^{\mathrm{magnon}}-\Gamma_{0}$, $\omega_{0}=\omega_{1}^{\mathrm{magnon}}$
and $\omega_{1}=\omega_{0}^{\mathrm{magnon}}$. We took our conventions
for the spin connection from \cite{Nicolai:1992xx}. Functions $p,q,r,s$ are
useful in what follows. %
}\begin{align*}
\rho_{0} & =\Gamma_{0}+c\gamma^{2}\frac{\cos^{2}\theta}{\sin\theta}\Gamma_{\phi}+\gamma^{2}\frac{\cos\theta}{\sin\theta}\sqrt{\sin^{2}\theta-c^{2}}\Gamma_{\theta}\qquad=\Gamma_{0}+r(\theta)\Gamma_{\phi}+s(\theta)\Gamma_{\theta}\,,\\
\rho_{1} & =\gamma^{2}\frac{\sin^{2}\theta-c^{2}}{\sin\theta}\Gamma_{\phi}-c\gamma^{2}\frac{\cos\theta}{\sin\theta}\sqrt{\sin^{2}\theta-c^{2}}\Gamma_{\theta}\qquad\:=\: p(\theta)\Gamma_{\phi}+q(\theta)\Gamma_{\theta}\,,\\
\omega_{0} & =-\omega_{0}^{\phi\theta}=-c\gamma^{2}\frac{\cos^{3}\theta}{\sin^{2}\theta}\,,\\
\omega_{1} & =-\omega_{1}^{\phi\theta}=-\gamma^{2}\frac{\cos\theta}{\sin^{2}\theta}(\sin^{2}\theta-c^{2})\,.\end{align*}

The first step is to replace $\partial_{0}=\partial_{t}$ and $\partial_{1}=\partial_{x}$
with the boosted derivatives $\partial_{u}=\gamma(\partial_{1}+c\partial_{0})$
and $\partial_{v}=\gamma(\partial_{0}+c\partial_{1})$, thus \[
\partial_{0}\pm\partial_{1}=(1\mp c)\gamma\left\{ \partial_{u}\pm\partial_{v}\right\} \,.\]
Following this pattern, define $G$ and $\tilde{G}$ as follows: \begin{align*}
\omega_{0}^{\phi\theta}+\omega_{1}^{\phi\theta} & =(1-c)\gamma\frac{1}{2}G, & \mbox{where}\quad G & =\gamma\frac{\cos\theta}{\sin^{2}\theta}(c+\sin^{2}\theta),\\
\omega_{0}^{\phi\theta}-\omega_{1}^{\phi\theta} & =(1+c)\gamma\frac{1}{2}\tilde{G}, & \tilde{G} & =\gamma\frac{\cos\theta}{\sin^{2}\theta}(c-\sin^{2}\theta).\end{align*}
 The equations of motion can then be written as\begin{align}
(\rho_{0}-\rho_{1})\left[(1-c)\gamma\left\{ \partial_{v}+\partial_{u}+\frac{1}{2}G\Gamma_{\phi\theta}\right\} \Theta^{1}-\frac{i}{2}\Gamma_{\star}(\rho_{0}+\rho_{1})\Theta^{2}\right] & =0,\label{eq:EQM-with-G}\\
(\rho_{0}+\rho_{1})\left[(1+c)\gamma\left\{ \partial_{v}-\partial_{u}+\frac{1}{2}\tilde{G}\Gamma_{\phi\theta}\right\} \Theta^{2}+\frac{i}{2}\Gamma_{\star}(\rho_{0}-\rho_{1})\Theta^{1}\right] & =0.\nonumber \end{align}
It is useful to define operators \begin{align*}
\mathcal{D}_{v} & =\partial_{v}+\frac{1}{2}G\Gamma_{\phi\theta}\;,\qquad\quad\tilde{\mathcal{D}}_{v}=\partial_{v}+\frac{1}{2}\tilde{G}\Gamma_{\phi\theta}\,,\end{align*}
so that the curly brackets in the equations of motion \eqref{eq:EQM-with-G}
are these operators plus or minus the time derivative $\partial_{u}$.

We now want to write the equations of motion in terms of kappa-symmetry
fixed fields, \cite{Green:1987sp}\cite{Minahan:2007gf,Papathanasiou:2007gd}
which we define as \begin{align}
\Psi^{1} & =-i(\rho_{0}-\rho_{1})\Theta^{1}\,,\label{Def of kappa-fixed Psis}\\
\Psi^{2} & =i(\rho_{0}+\rho_{1})\Theta^{2}\,.\nonumber \end{align}
Note that $\Gamma_{11}$ anti-commutes with $i(\rho_{0}\pm\rho_{1})$,
and that these operators are real. Thus $\Theta^{I}$ is Majorana--Weyl
exactly when $\Psi^{I}$ is, so we will impose the conditions on $\Psi^{I}$.

To write the equations of motion in terms of these symmetry-fixed
fields, we will need several identities, identical in form to those
in the giant magnon case \cite{Minahan:2007gf}.%
\footnote{To derive these identities, write relations such as $(\rho_{0}\pm\rho_{1})^{2}=-1+(r\pm p)^{2}+(s\pm q)^{2}$
and $(\overline{\rho}_{0}\pm\rho_{1})^{2}=-1+(-r\pm p)^{2}+(-s\pm q)^{2}$. %
} The following two operators are nilpotent:\[
\left(\rho_{0}\pm\rho_{1}\right)^{2}=0\]
(thus $(\rho_{0}-\rho_{1})\Psi^{1}=0$ and $(\rho_{0}+\rho_{1})\Psi^{2}=0$)
and can be shown to commute with the curly derivatives\begin{align*}
\left[\mathcal{D}_{v},(\rho_{0}-\rho_{1})\right] & =0\;,\qquad\quad\left[\tilde{\mathcal{D}}_{v},(\rho_{0}+\rho_{1})\right]=0\,.\end{align*}
(They trivially commute with $\partial_{u}$ too.) Also important
is the dagger of $\rho_{0}$:\[
\overline{\rho}_{0}\equiv\Gamma_{\star}\rho_{0}\Gamma_{\star}=-\rho_{0}^{\dagger}=\Gamma_{0}-r\Gamma_{\phi}-s\Gamma_{\theta},\]
which allows us to write two more nilpotent operators\[
\left(\overline{\rho}_{0}\pm\rho_{1}\right)^{2}=0\]
as well as a non-singular operator $(\overline{\rho}_{0}-\rho_{0})=-2r\Gamma_{\phi}-2s\Gamma_{\theta}$,
whose square is proportional to the unit matrix: \begin{align*}
\left(\overline{\rho}_{0}-\rho_{0}\right)^{2} & =4\gamma^{2}\cos^{2}\theta.\end{align*}

Returning to the equations of motion \eqref{eq:EQM-with-G}, we can
now pull the operators $(\rho_{0}\pm\rho_{1})$ to the right, using
the identities above, until they act on the $\Theta^{I}$ to give
$\Psi^{I}$. We obtain: \begin{align}
(1-c)\gamma\left\{ \mathcal{D}_{v}+\partial_{u}\right\} \Psi^{1}+\frac{i}{2}\Gamma_{\star}(\overline{\rho}_{0}+\rho_{0})\Psi^{2} & =0\,,\label{eq:Equations of motion Psi}\\
(1+c)\gamma\left\{ \tilde{\mathcal{D}}_{v}-\partial_{u}\right\} \Psi^{2}-\frac{i}{2}\Gamma_{\star}(\overline{\rho}_{0}-\rho_{0})\Psi^{1} & =0\,.\nonumber \end{align}

\subsection{Non-zero modes\label{sub:fermi-nonzero-modes}}

Begin by solving the first of equations \eqref{eq:Equations of motion Psi}
for $\Psi^{2}$ :\begin{equation}
\Psi^{2}=\frac{(\overline{\rho}_{0}-\rho_{0})}{4\gamma^{2}\cos^{2}\theta}\Gamma_{\star}\frac{2}{i}(1-c)\gamma\left\{ \mathcal{D}_{v}+\partial_{u}\right\} \Psi^{1}.\label{Psi2 in terms of Psi1}\end{equation}
We can then eliminate $\Psi^{2}$ from the other equation to obtain
a second-order equation for $\Psi^{1}$ alone: \[
\left\{ \tilde{\mathcal{D}}_{v}-\partial_{u}\right\} \frac{(\overline{\rho}_{0}-\rho_{0})}{\gamma^{2}\cos^{2}\theta}\left\{ \mathcal{D}_{v}+\partial_{u}\right\} \Psi^{1}+(\overline{\rho}_{0}-\rho_{0})\Psi^{1}=0.\]
Using the identity\[
\left\{ \tilde{\mathcal{D}}_{v}-\partial_{u}\right\} \frac{(\overline{\rho}_{0}-\rho_{0})}{\cos\theta}=\frac{(\overline{\rho}_{0}-\rho_{0})}{\cos\theta}\left\{ \mathcal{D}_{v}-\partial_{u}\right\} \]
and pulling the $(\rho_{0}-\rho_{1})$ from $\Psi^{1}$'s definition
through, this becomes \begin{equation}
(\rho_{0}-\rho_{1})\left(\frac{1}{\gamma\cos\theta}\left\{ \mathcal{D}_{v}-\partial_{u}\right\} \frac{1}{\gamma\cos\theta}\left\{ \mathcal{D}_{v}+\partial_{u}\right\} +1\right)\Theta^{1}=0\,,\label{eq:second-order Theta eq}\end{equation}
analogous to equation (3.7) of \cite{Papathanasiou:2007gd}. We can
solve this equation by a similar method to the one used there: we
temporarily drop the kappa-symmetry projection $(\rho_{0}-\rho_{1})$,
and solve the remainder of the equation for $\Theta^{1}$. At the
end we will apply the projection to recover $\Psi^{1}$, and from
that will find $\Psi^{2}$ using \eqref{Psi2 in terms of Psi1}. 

To find $\Theta^{1}$, we split this second-order equation \eqref{eq:second-order Theta eq}
into two first-order equations, defining some intermediate field $\tilde{\Theta}$:\[
\left[\begin{array}{cc}
\mathcal{D}_{v}+\partial_{u} & -i\sech v\\
-i\sech v & \mathcal{D}_{v}-\partial_{u}\end{array}\right]\left(\begin{array}{c}
\Theta^{1}\\
\tilde{\Theta}^{}\end{array}\right)=0\,.\]
We expand the spinor in a Fourier series for $u$:\begin{equation}
\left(\begin{array}{c}
\Theta^{1}\left(u,v\right)\\
\tilde{\Theta}\left(u,v\right)\end{array}\right)=e^{-i\omega u}\vec{\Theta}(v,\omega)\,,\label{eq:Fourier expansion}\end{equation}
and also into a sum of eigenspinors of $\Gamma_{\phi\theta}$: $\vec{\Theta}=\vec{\Theta}_{+}+\vec{\Theta}_{-}$
with\[
\left(1_{2\times2}\otimes\Gamma_{\phi\theta}\right)\:\vec{\Theta}_{\pm}=\pm i\,\vec{\Theta}_{\pm}.\]
Then the coupled linear equations can be written as\[
\left(\partial_{v}-V_{\pm}\right)\vec{\Theta}_{\pm}=0,\qquad\mbox{with}\quad V_{\pm}=\left[\begin{array}{cc}
i\left(\omega\mp\frac{G}{2}\right) & i\sech v\\
i\sech v & -i\left(\omega\pm\frac{G}{2}\right)\end{array}\right].\]
We now proceed to diagonalize the system of equations, by a change
of basis.

\subsubsection{Diagonalisation}

Define $\vec{\Theta}'_{\pm}=S\vec{\Theta}_{\pm}$, which obeys \begin{align}
\partial_{v}\vec{\Theta}'_{\pm} & =(\partial_{v}S+SV_{\pm})S^{-1}\vec{\Theta}'_{\pm}=H_{\pm}\vec{\Theta}'_{\pm}\label{eq:EOM diagonalized}\end{align}
 (defining $H_{\pm}$). We want to choose $S$ to make $H_{\pm}$
diagonal. If we write\[
S=\left[\begin{array}{cc}
a(v) & b(v)\\
c(v) & d(v)\end{array}\right],\]
then setting the off-diagonal elements of $H_{\pm}$ to zero reads
\begin{align*}
0 & =i(a^{2}-b^{2})\sech v-2i\omega ab-ba'+ab',\\
0 & =i(d^{2}-c^{2})\sech v+2i\omega cd-cd'+dc'.\end{align*}
One obtains the same equations for both $H_{+}$ and $H_{-}$, which
means $S$ diagonalizes both simultaneously. Because we have only
two equations and four parameters, we choose two additional relations
among these four entries%
\footnote{These extra relations can be imposed by multiplying $S$ by a non-singular
diagonal matrix, which is always allowed as it does not change the
equations of motion \eqref{eq:EOM diagonalized}.%
}\begin{align}
a' & =-ib\sech(v)\,,\label{eq:extra-rel-ab-cd}\\
c' & =-id\sech(v)\,,\nonumber \end{align}
leading to a second-order equation for $a$ (and $c$ obeying the
same equation):\begin{equation}
-a''-\tanh v\, a'+2i\omega a'-\sech^{2}v\, a=0.\label{eq:2nd order eqn for entry a}\end{equation}
It also leads to this simple form for $H_{\pm}$:\begin{equation}
H_{\pm}=i\left(\omega\mp\frac{G}{2}\right)\left[\begin{array}{cc}
1 & 0\\
0 & 1\end{array}\right]\,.\label{eq:The "hamiltonian"}\end{equation}

The two solutions to \eqref{eq:2nd order eqn for entry a} are \begin{align*}
a_{1}\left(v\right) & =\frac{2i\omega}{1+4\omega^{2}}+\frac{\tanh v}{1+4\omega^{2}}\,,\\
a_{2}\left(v\right) & =e^{2i\omega v}\sech v\,,\end{align*}
and $a(v)$ and $c(v)$ are (different) linear combinations of these.
The other functions $b(v)$ and $d(v)$ are then fixed by \eqref{eq:extra-rel-ab-cd}.
We can write the general solution for $S$ as \[
S=S_{0}\left[\begin{array}{cc}
a_{1}(v) & b_{1}(v)\\
a_{2}(v) & b_{2}(v)\end{array}\right]\]
where $S_{0}$ is a non-singular constant matrix, and \begin{align*}
b_{1}\left(v\right) & =i\frac{\sech v}{1+4\omega^{2}}\,,\\
b_{2}\left(v\right) & =i\left(2i\omega-\tanh v\right)e^{2i\omega v}\,.\end{align*}

The determinant of this change of basis is $\det S=-ie^{2i\omega v}\,\det S_{0}$,
different from zero, as expected.

\subsubsection{Solving}

We can now solve the diagonalized system \eqref{eq:EOM diagonalized},
using $H_{\pm}$ from \eqref{eq:The "hamiltonian"}. The equations
become simply \[
\left(\partial_{v}-i\left(\omega\mp\frac{G}{2}\right)\right)f(v)=0\,.\]
A very similar equation occurs in the magnon case \cite{Minahan:2007gf}
(and also for the zero modes in appendix \ref{sec:Fermionic-Zero-modes}).
It has solution $f(v)=e^{\pm i\chi}e^{i\omega v}$, where \[
e^{i\chi}=\left(\frac{\sinh v+ic}{\sinh v-ic}\right)^{1/4}\sqrt{\tanh v+i\sech v}.\]
 Thus $\vec{\Theta}'_{\pm}$ will be given by this phase times a spinor:
\[
\vec{\Theta}'_{\pm}=e^{\pm i\chi}e^{i\omega v}\vec{U}_{\pm}\,,\]
where $\vec{U}_{\pm}$ is any eigenspinor of $\left(1\otimes\Gamma_{\phi\theta}\right)$
with eigenvalues $\pm i$. It remains to rotate back to unprimed $\vec{\Theta}_{\pm}$,
which is\[
\vec{\Theta}_{\pm}=S^{-1}\vec{\Theta}_{\pm}=e^{\pm i\chi}e^{i\omega v}S^{-1}\vec{U}_{\pm}.\]
We can now absorb the constant matrix $S_{0}^{-1}$ into the arbitrary
spinor $\vec{U}_{\pm}$, which we do by writing\[
S_{0}^{-1}\vec{U}_{\pm}=\frac{1}{\sqrt{1-c}}\left(\begin{array}{c}
U_{\pm}\\
\tilde{U}_{\pm}\end{array}\right)\,,\]
introducing $U_{\pm}$ and $\tilde{U}_{\pm}$, and slipping in the
$\sqrt{1-c}$ for later convenience.

Recall from \eqref{eq:Fourier expansion} that our original spinor
$\Theta^{1}$ is the first component of $e^{-i\omega u}\left(\vec{\Theta}_{+}+\vec{\Theta}_{-}\right)$.
We can now read it off, obtaining%
\footnote{The second entry will be given by\[
\tilde{\Theta}_{}\left(u,v\right)=-ie^{-i\omega u}e^{\pm i\chi}\left[e^{i\omega v}\sech v\, U_{\pm}-\frac{\left(\tanh v+2i\omega\right)}{1+4\omega^{2}}e^{-i\omega v}\tilde{U}_{\pm}\right].\]
This is useful for finding $\Psi^{2}$ later.%
}\begin{align*}
\Theta^{1}\left(u,v\right) & =\frac{1}{\sqrt{1-c}}e^{-i\omega u}\sum_{\pm}\frac{e^{i\omega v\pm i\chi}}{-ie^{2i\omega v}}\left[b_{2}\left(v\right)U_{\pm}-b_{1}\left(v\right)\tilde{U}_{\pm}\right]\\
 & =\frac{-1}{\sqrt{1-c}}e^{-i\omega u}\sum_{\pm}e^{\pm i\chi}\left[e^{-i\omega v}\frac{\sech v}{1+4\omega^{2}}U_{\pm}+\left(\tanh v-2i\omega\right)e^{i\omega v}\tilde{U}_{\pm}\right].\end{align*}
 To get the symmetry-fixed field $\Psi^{1}=-i(\rho_{0}-\rho_{1})\Theta^{1}$
it is useful to use the identity $e^{\pm2i\chi}=(p-r)\mp i(q-s)$.
We find the following positive-frequency solution:\begin{align}
\Psi_{p}^{1} & =\frac{i\, e^{-i\omega u}}{\sqrt{1-c}}\sum_{\pm}\left(e^{\pm i\chi}\Gamma_{0}-e^{\mp i\chi}\Gamma_{\phi}\right)\left[e^{-i\omega v}\frac{\sech v}{1+4\omega^{2}}U_{\pm}^{}+\left(\tanh v-2i\omega\right)e^{i\omega v}\tilde{U}_{\pm}\right]\nonumber \\
 & =\frac{i}{\sqrt{1-c}}\sum_{\pm}\left(e^{\pm i\chi}\Gamma_{0}-e^{\mp i\chi}\Gamma_{\phi}\right)\left[e^{i\alpha}\frac{\sech v}{1+4\omega^{2}}U_{\pm}^{}+\sqrt{\tanh^{2}v+4\omega^{2}}e^{i\beta}\tilde{U}_{\pm}\right]\label{eq:Psi1 pre-majorana}\end{align}
where the phases $\alpha$ and $\beta$ are defined by\begin{eqnarray*}
e^{i\alpha} & = & e^{-i\omega\left(u+v\right)},\\
e^{i\beta} & = & e^{-i\omega\left(u-v\right)}e^{-i\arctan\left(2\omega\coth v\right)}.\end{eqnarray*}

\subsubsection{Majorana condition}

It remains to impose the Majorana condition on the spinors, that is,
$\Psi^{I}$ should be real $\Psi^{I*}=\Psi^{I}$. To do so, we have
to consider a superposition of positive and negative frequencies $\omega$.
We thus write \begin{align*}
\Psi^{1} & =2\re\Psi_{p}^{1}=\Psi_{p}^{1}+\Psi_{p}^{1*}\\
 & =\frac{i}{\sqrt{1-c}}\sum_{\pm}\left(e^{\pm i\chi}\Gamma_{0}-e^{\mp i\chi}\Gamma_{\phi}\right)\left[\frac{\sech v}{1+4\omega^{2}}\left(e^{i\alpha}U_{\pm}+e^{-i\alpha}U_{\mp}^{*}\right)\right.\\
 & \qquad\qquad\qquad\qquad\quad\left.+\sqrt{\tanh^{2}v+4\omega^{2}}\left(e^{i\beta}\tilde{U}_{\pm}+e^{-i\beta}\tilde{U}_{\mp}^{*}\right)\right].\end{align*}
Note that $U_{\mp}^{*}$ is an eigenspinor of $\Gamma_{\phi\theta}$
of eigenvalue $\pm i$. (The $\Gamma$ matrices are imaginary, thus
$\Gamma_{\phi\theta}$ is real.) 

Combine the four $\pm$ eigenspinors into two spinors $U=U_{+}+U_{-}$
and $\tilde{U}=\tilde{U}_{+}+\tilde{U}_{-}$. (We can reverse this
with projection operators $U_{\pm}=\frac{i\pm\Gamma_{\phi\theta}}{2i}U$,
and similarly for the others.). Then we can write \begin{align}
\Psi^{1} & =\frac{i}{\sqrt{1-c}}\left[\Gamma_{0}\left(\cos\chi+\Gamma_{\phi\theta}\sin\chi\right)-\Gamma_{\phi}\left(\cos\chi-\Gamma_{\phi\theta}\sin\chi\right)\right]\nonumber \\
 & \qquad\qquad\times\left\{ \frac{\sech v}{1+4\omega^{2}}\re(e^{i\alpha}U)+\sqrt{\tanh^{2}v+4\omega^{2}}\re(e^{i\beta}\tilde{U})\right\} \nonumber \\
 & =\frac{i}{\sqrt{1-c}}\left[\Gamma_{0}\left(\cos\chi+\Gamma_{\phi\theta}\sin\chi\right)-\Gamma_{\phi}\left(\cos\chi-\Gamma_{\phi\theta}\sin\chi\right)\right]\nonumber \\
 & \qquad\qquad\times\Big\lbrace\frac{\sech v}{1+4\omega^{2}}\left(\cos\alpha\, U_{0}+\sin\alpha\,\Gamma_{\phi\theta}U_{1}\right)\nonumber \\
 & \qquad\qquad\qquad+\sqrt{\tanh^{2}v+4\omega^{2}}\left(\cos\beta\,\tilde{U}_{0}+\sin\beta\,\Gamma_{\phi\theta}\tilde{U}_{1}\right)\Big\rbrace,\label{eq:nonzero psi1}\end{align}
where the new spinors are\begin{align}
U_{0} & =2\re\left(U_{+}+U_{-}\right)\,, & \tilde{U}_{0} & =2\re\left(\tilde{U}_{+}+\tilde{U}_{-}\right)\,,\label{eq:U0U1-real-Upm}\\
U_{1} & =2\re\left(U_{+}-U_{-}\right)\,, & \tilde{U}_{1} & =2\re\left(\tilde{U}_{+}-\tilde{U}_{-}\right)\,,\nonumber \end{align}
thus $U_{0}=2\re U$, but $U_{1}=2\Gamma_{\phi\theta}\im U$ (and
similarly with tildes).

We can now find $\Psi^{2}$ from $\Psi^{1}$ using \eqref{Psi2 in terms of Psi1}.
The final, Majorana, field is\begin{align}
\Psi^{2} & =\frac{1}{\sqrt{1+c}}\Gamma_{*}\Gamma_{\theta}\left[\Gamma_{0}\left(\cos\tilde{\chi}+\Gamma_{\phi\theta}\sin\tilde{\chi}\right)-\Gamma_{\phi}\left(\cos\tilde{\chi}-\Gamma_{\phi\theta}\sin\tilde{\chi}\right)\right]\nonumber \\
 & \qquad\qquad\times\Big\lbrace\sech v\left(\cos\tilde{\alpha}\,\tilde{U}_{0}+\sin\tilde{\alpha}\,\Gamma_{\phi\theta}\tilde{U}_{1}\right)\nonumber \\
 & \qquad\qquad\qquad-\frac{\sqrt{\tanh^{2}v+4\omega^{2}}}{1+4\omega^{2}}\left(\cos\tilde{\beta}\, U_{0}+\sin\tilde{\beta}\,\Gamma_{\phi\theta}U_{1}\right)\Big\rbrace,\label{eq:nonzero psi2}\end{align}
where the new phases are\begin{align*}
e^{i\tilde{\chi}} & =\sqrt{\frac{\sinh v-ic}{\sinh v+ic}}e^{i\chi}=\left(\frac{\sinh v-ic}{\sinh v+ic}\right)^{1/4}\sqrt{\tanh v+i\sech v},\\
e^{i\tilde{\alpha}} & =e^{-i\omega\left(u-v\right)},\\
e^{i\tilde{\beta}} & =e^{-i\omega\left(u+v\right)}e^{i\arctan\left(2\omega\coth v\right)}.\end{align*}

\subsection{Mass and Counting}

The phases appearing in $\Psi^{1}$, far from the spike ($\left|v\right|\gg1$),
are $i\alpha=-i\omega u-i\omega v$ and $i\beta=-i\omega u+i\omega v$.
This means that the fermionic modes are massless, $\omega^{2}=k^{2}$.
But the bosonic modes we found in section \ref{sec:Bosonic-Sector}
are not massless, so there can be no supersymmetry.

How many fermionic modes are there? There are four spinors $U_{\pm}$
and $\tilde{U}_{\pm}$, which are $\Gamma_{\phi\theta}$ eigenspinors,
so have 16 complex components each. They must also be $\Gamma_{11}$
eigenspinors, for $\Psi$ to be Weyl, cutting the number by half.
And then we found in \prettyref{eq:U0U1-real-Upm} that the Majorana
spinor depends only on the real part of each, cutting it in half again.
This leaves 16 complex degrees of freedom, which is twice the number
for the giant magnon. \cite{Papathanasiou:2007gd}

The bosonic modes were obtained simply by switching $x\leftrightarrow t$
in the magnon expressions. So their number is unchanged from the magnon
case: there are 8 non-zero modes (4 on the sphere and 4 in $AdS$).
The fact that there are two fermionic modes for each bosonic modes
is a second piece of evidence against supersymmetry.

There are also twice as many fermionic zero modes (8 complex) as bosonic
zero modes (4, as for the magnon). Because the non-zero modes are
massless, $\omega=0$ is part of the continuum, and expressions for
the zero modes can be found by simply setting $\omega=0$ in \eqref{eq:nonzero psi1}
and \eqref{eq:nonzero psi2} above (which sets $\alpha=\beta=0$).
But the counting is more delicate, the zero modes appear to have the
same dependence on $\re U_{\pm}$ and $\re\tilde{U}_{\pm}$ as the
non-zero modes, suggesting that there are also 16 of them. However,
the same argument as used for the magnon case \cite{Minahan:2007gf}
kills half of these, leaving 8. We perform the exact analogue of \cite{Minahan:2007gf}'s
calculation of the magnon zero modes in appendix \ref{sec:Fermionic-Zero-modes}.

\section{Quantum Corrections\label{sec:Section-Four}}

\subsection{Corrections to what?\label{sub:Corrections-to-what?}}

Having found the modes, it would be natural to use them to compute
a first quantum correction, i.e. to perform `semi-classical quantisation'.
For the giant magnon, this means finding quantum corrections to $\Delta-J$.
The origin of this is as follows:

Frolov and Tseytlin \cite{Frolov:2002av} consider the `vacuum' of
the large-$J$ sector, the point particle orbiting the sphere, which
has $\Delta=J$. They add small perturbations to this, and show that
$\Delta-J$ is (at leading order in $1/\sqrt{\lambda}$) the Hamiltonian
of a 1+1-dimensional theory. The perpendicular fluctuations in both
the sphere and $AdS$ are non-interacting massive fields of this theory.
So far this is classical. The semi-classical correction is to treat
each mode of these fields as a harmonic oscillator, and their zero-point
energies `$\frac{1}{2}\hbar\omega$' are corrections to $\Delta-J$.
The magnon is interpreted as a `giant perturbation' of this vacuum,
tall enough to see the curvature of spacetime. (And, it turns out,
of high enough momentum to see that the 1+1-dimensional theory is
a spin chain, with periodic dispersion relation.)

Here we repeat their calculation, for the `vacuum of the large-winding
sector': the infinitely wound hoop. We find as Hamiltonian $\Delta-\Phi$,
with the winding charge $\Phi$ replacing the angular momentum $J$.
The single spike is similarly a `giant perturbation' of this vacuum.

Recall from section \ref{sub:flat-space} that the flat-space versions
of these two classes of spiky strings are related by T-duality, which
famously exchanges winding and momentum around a compact direction.
Clearly this change in the Hamiltonian is somehow a consequence of
this duality. But notice that the compact direction here is part of
a sphere, and that the radius of this sphere is unchanged.

\subsubsection{Finding the Hamiltonian}

Write the metric in the form%
\footnote{The azimuthal angle $\phi$ here is the same as used before, in \eqref{eq:S2-metric},
but $\theta_{4}=\pi/2-\theta$ is the elevation above the equator.
The expansions of the metric components which we need are $G_{\tau\tau}=-1-\eta^{2}+\cdots$
and $G_{\theta\theta}=1-\theta^{2}+\cdots$. %
}\begin{align*}
ds_{\mathrm{AdS}}^{2} & =-\left(\frac{1+\eta^{2}/4}{1-\eta^{2}/4}\right)^{2}d\tau^{2}+\frac{1}{\left(1-\eta^{2}/4\right)^{2}}d\eta_{k}d\eta_{k}\qquad\quad k=1,2,3,4\,\\
ds_{\mathrm{S}}^{2} & =d\theta_{1}^{2}+\cos^{2}\theta_{1}\left(d\theta_{2}^{2}+\cos^{2}\theta_{2}\left(d\theta_{2}^{2}+\cos^{2}\theta_{2}\left(d\theta_{3}^{2}+\cos^{2}\theta_{3}\left(d\theta_{4}^{2}+\cos^{2}\theta_{4}d\phi^{2}\right)\right)\right)\right)\,.\end{align*}

The action (in conformal gauge) is \begin{equation}
S=-\frac{\sqrt{\lambda}}{2\pi}\int dxdt\:\mathcal{L}_{B},\qquad\qquad\mathcal{L}_{B}=\frac{1}{2}\partial^{a}X^{\mu}\partial_{a}X^{\nu}G_{\mu\nu}\,.\label{eq:string lagrang}\end{equation}
We write the perturbed the solution as $X^{\mu}=X_{\mathrm{hoop}}^{\mu}+\tilde{X}^{\mu}/\lambda^{1/4}$:\begin{align}
\tau & =t+\frac{1}{\lambda^{1/4}}\tilde{\tau}\, & \phi & =x+\frac{1}{\lambda^{1/4}}\tilde{\phi}\,\label{eq:expansion in fluct}\\
\eta_{k} & =\frac{1}{\lambda^{1/4}}\tilde{\eta}_{k}\, & \theta_{s} & =\frac{1}{\lambda^{1/4}}\tilde{\theta}_{s},\qquad s=1,2,3,4.\nonumber \end{align}
Expanding at large $\lambda$, the Lagrangian becomes\begin{align}
\mathcal{L}_{B} & =1+\frac{1}{\lambda^{1/4}}\left(\partial_{0}\tilde{\tau}+\partial_{1}\tilde{\phi}\right)\nonumber \\
 & \quad+\frac{1}{2\sqrt{\lambda}}\left(-\partial^{a}\tilde{\tau}\partial_{a}\tilde{\tau}+\partial^{a}\tilde{\eta}_{k}\partial_{a}\tilde{\eta}_{k}+\partial^{a}\tilde{\phi}\partial_{a}\tilde{\phi}+\partial^{a}\tilde{\theta}_{s}\partial_{a}\tilde{\theta}_{s}+\tilde{\eta}_{k}\tilde{\eta}_{k}-\tilde{\theta}_{s}\tilde{\theta}_{s}\right)\nonumber \\
 & \quad+\frac{1}{\lambda^{3/4}}\left((\partial_{0}\tilde{\tau})\tilde{\eta}_{k}\tilde{\eta}_{k}-(\partial_{1}\tilde{\phi})\tilde{\theta}_{s}\tilde{\theta}_{s}\right)+\mathcal{O}(\frac{1}{\lambda}).\label{eq:Lagr fluctuations}\end{align}
In the quadratic piece, $\tilde{\eta}_{k}$ appears massive and $\tilde{\theta}_{s}$
tachyonic, matching what we found for the single spike's modes.

The Virasoro constraints are first $\gamma_{00}+\gamma_{11}=2T_{00}=0$:\begin{align}
0 & =\frac{1}{\lambda^{1/4}}\left(-\partial_{0}\tilde{\tau}+\partial_{1}\tilde{\phi}\right)\nonumber \\
 & \quad+\frac{1}{2\sqrt{\lambda}}\left(-\partial_{a}\tilde{\tau}\partial_{a}\tilde{\tau}+\partial_{a}\tilde{\eta}_{k}\partial_{a}\tilde{\eta}_{k}+\partial_{a}\tilde{\phi}\partial_{a}\tilde{\phi}+\partial_{a}\tilde{\theta}_{s}\partial_{a}\tilde{\theta}_{s}-\tilde{\eta}_{k}\tilde{\eta}_{k}-\tilde{\theta}_{s}\tilde{\theta}_{s}\right)\nonumber \\
 & \quad+\mathcal{O}(\frac{1}{\lambda^{3/4}}),\label{eq:Virasoro const}\end{align}
(writing $\partial_{a}\partial_{a}=\partial_{0}\partial_{0}+\partial_{1}\partial_{1}$
in a temporary abuse of notation) and second $\gamma_{01}=T_{01}=0$:\begin{align*}
0 & \!=\frac{1}{\lambda^{1/4}}\left(-\partial_{1}\tilde{\tau}+\partial_{0}\tilde{\phi}\right)+\frac{1}{2\sqrt{\lambda}}\left(-\partial_{0}\tilde{\tau}\partial_{1}\tilde{\tau}+\partial_{0}\tilde{\phi}\partial_{1}\tilde{\phi}+\partial_{0}\tilde{\eta}_{k}\partial_{1}\tilde{\eta}_{k}+\partial_{0}\tilde{\theta}_{s}\partial_{1}\tilde{\theta}_{s}\right)\!+\mathcal{O}(\frac{1}{\lambda^{3/4}}).\end{align*}

Now we expand the spacetime charges: the energy is the integral of
the momentum density $\Pi_{\tau}^{0}$:\begin{align*}
\Delta & =\frac{1}{2\pi}\int dx\:\frac{\partial\mathcal{L}_{B}}{\partial\,\partial_{0}\tau}\\
 & =\frac{1}{2\pi}\int dx\left(\sqrt{\lambda}+\lambda^{1/4}\partial_{0}\tilde{\tau}+\tilde{\eta}_{k}\tilde{\eta}_{k}+\mathcal{O}(\frac{1}{\lambda^{1/4}})\right)\,,\end{align*}
and the winding charge defined in \eqref{eq:winding-defn} is \begin{align*}
\Phi & =\frac{\sqrt{\lambda}}{2\pi}\int dx\:\partial_{1}\phi\\
 & =\frac{1}{2\pi}\int dx\left(\sqrt{\lambda}+\lambda^{1/4}\partial_{1}\tilde{\phi}\right)\,.\end{align*}
Subtracting these two charges, the two $\sqrt{\lambda}$ terms will
cancel, leaving a finite result. The linear terms can then be replaced
with quadratic terms using the first Virasoro constraint \eqref{eq:Virasoro const}.
To leading order in $1/\lambda$, we obtain:\begin{align}
\Delta-\Phi & =\frac{1}{4\pi}\int dx\bigg[-\left(\partial_{0}\tilde{\tau}\partial_{0}\tilde{\tau}+\partial_{1}\tilde{\tau}\partial_{1}\tilde{\tau}\right)+\left(\partial_{0}\tilde{\phi}\partial_{0}\tilde{\phi}+\partial_{1}\tilde{\phi}\partial_{1}\tilde{\phi}\right)\nonumber \\
 & \qquad+\left(\partial_{0}\tilde{\eta}_{k}\partial_{0}\tilde{\eta}_{k}+\partial_{1}\tilde{\eta}_{k}\partial_{1}\tilde{\eta}_{k}\right)+\left(\partial_{0}\tilde{\theta}_{s}\partial_{0}\tilde{\theta}_{s}+\partial_{1}\tilde{\theta}_{s}\partial_{1}\tilde{\theta}_{s}\right)+\tilde{\eta}_{k}\tilde{\eta}_{k}-\tilde{\theta}_{s}\tilde{\theta}_{s}\bigg].\label{eq:DeltaPhi-hamilt}\end{align}
This is the analogue of the result in \cite{Frolov:2002av}. The fields
$\tilde{\tau}$ and $\tilde{\phi}$ correspond to transformations
that are pure gauge, so we drop them. We can write $\Delta-\Phi$
in terms of the Hamiltonian one would obtain from only the quadratic
part of the Lagrangian $\mathcal{L}_{B}$ (see appendix \prettyref{sub:2-dimensional-Hamiltonian}),
which contains the transverse (physical) modes $\tilde{\eta}_{k}$
and $\tilde{\theta}_{s}$ and their conjugate momenta $\tilde{\Pi}_{\tilde{\eta}_{k}},\tilde{\Pi}_{\tilde{\theta}_{s}}$:
\begin{align*}
\Delta-\Phi & =\int\frac{dx}{2\pi}\mathcal{H}_{2d}\left(\tilde{\tau},\tilde{\phi},\tilde{\eta}_{k},\tilde{\theta}_{s}\right)\\
 & =\sqrt{\lambda}\int\frac{dx}{4\pi}\left[\tilde{\Pi}_{\tilde{\eta}_{k}}^{2}+\tilde{\Pi}_{\tilde{\theta}_{s}}^{2}+\partial_{1}\tilde{\eta}_{k}\partial_{1}\tilde{\eta}_{k}+\partial_{1}\tilde{\theta}_{s}\partial_{1}\tilde{\theta}_{s}+\tilde{\eta}_{k}\tilde{\eta}_{k}-\tilde{\theta}_{s}\tilde{\theta}_{s}\right].\end{align*}

We are left with four massive fields from vibrations in the $AdS$
directions and four tachyonic fields from the sphere directions. Then
$\Delta-\Phi$ is the expected quadratic Hamiltonian for these 8 fields.
One could perform a similar construction for the fermionic modes obtaining
16 massless fermionic fields. \cite{Metsaev:2001bj,Metsaev:2002re}\cite{Frolov:2002av}

\subsection{First quantum correction}

For each of the eight bosonic modes $\tilde{\eta}_{k}$ and $\tilde{\theta}_{s}$,
we have a quadratic Hamiltonian of the kind\[
H_{2}=\int dx\left[\frac{1}{2}\hat{\Pi}^{2}+\hat{\phi}\left(-\partial_{x}^{2}+V\right)\hat{\phi}\right].\]
Note that $V=\pm1$ in our case, depending on whether the mode is
massive or tachyonic. We can expand both $\hat{\Pi}$ and $\hat{\phi}$
eigenfunctions $\psi_{n}$ of the differential operator $\left(-\partial_{x}^{2}+V\right)\psi_{n}=\omega_{n}^{2}\psi_{n}$,
which we write $\hat{\phi}=\sum\hat{\phi}_{n}\psi_{n}$ and $\hat{\Pi}=\sum\hat{\Pi}_{n}\psi_{n}$.
The Hamiltonian becomes a sum of decoupled harmonic oscillators\[
H_{2}=\sum\frac{1}{2}\left(\hat{\Pi}_{n}^{2}+\omega_{n}^{2}\hat{\phi}_{n}^{2}\right).\]
By introducing creation and annihilation operators in the usual way,
for each oscillator, we find that each of these contributes $\frac{1}{2}\sum\hbar\omega_{n}$,
with%
\footnote{In the literature, $\nu_{n}=T\omega_{n}$ (where $T$ is some large
time) is called a stability angle.%
} $\omega_{n}=\sqrt{k_{n}^{2}+m^{2}}$, for some mass $m^{2}$ and
allowed momenta $k_{n}$. 

For our solution the bosonic modes of section \ref{sec:Bosonic-Sector}
have $W(K)=\sqrt{K^{2}\pm1}$. Each of the fermionic modes will contribute
$-\frac{1}{2}\sum\hbar W_{\mathrm{fermi}}$, where the fermionic modes
found in section \ref{sec:Fermionic-Sector} have $W(K)=K$.

There are two important issues here:

\begin{itemize}
\item First, to obtain a finite first quantum correction for any solution,
one must always subtract the quantum correction for the corresponding
vacuum solution. Both of these are normally UV divergent (and this
subtraction is not the only renormalization usually needed). For the
single spike, the relevant vacuum is the hoop solution. Note that
the hoop has $\Delta-\Phi=0$ classically, so this subtraction is
only needed for the quantum corrections. 
\item Second, we are interested in studying those modes of the spike which
result in its instability. To determine the decay time of this unstable
solution, we are only interested in the imaginary part of the energy
correction. None of the fermionic modes will contribute to this, as
they are massless, nor will the 4 bosonic modes in $AdS_{5}$, as
they are massive. The only contribution is from the 4 tachyonic modes
on the sphere, which have $W(K)=\pm\sqrt{K^{2}-1}$, and here only
from those modes with $\left|K\right|<1$. This excludes the UV modes,
and in fact no other renormalization will be needed.
\end{itemize}

\subsubsection*{Vacuum }

The bosonic and fermionic modes for the hoop can be found in appendix
\ref{sub:hoop-modes}. They have the same masses as their counterparts
for the single spike, in particular the sphere modes have $W(K)=\pm\sqrt{K^{2}-1}$.
To discretise the momentum $K$, we put the solution in a box $-\frac{L}{2}<x<\frac{L}{2}$
and impose periodic boundary conditions $\delta X\left(-\frac{L}{2}\right)=\delta X\left(\frac{L}{2}\right)$.
Then $K_{n}=\frac{2\pi n}{L}$, with $n\in\mathbb{Z}$, and the contribution
of these modes to the vacuum energy is given by \begin{align}
\Delta E_{\mathrm{hoop}} & =4\frac{1}{2}\sum_{n}\sqrt{K_{n}^{2}-1}\nonumber \\
 & \approx2\frac{L}{2\pi}\int_{-1}^{1}dK\,\sqrt{K^{2}-1}\quad\mbox{as }L\rightarrow\infty\nonumber \\
 & =\frac{i}{2}L\,.\label{eq:DeltaE-vac}\end{align}
The integration is over $\left|K\right|<1$ because we are looking
for just the imaginary part. We do not encounter a UV divergence here.

\subsubsection*{Spike solution }

Again we study only the bosonic modes on the sphere with $\left|K\right|<1$.
But the discrete momenta $K$ allowed for the spike are not the same
as those for the hoop $K_{n}$, as the modes have a phase shift at
large $x$ compared to the hoop. Looking at the bosonic sphere modes
given in \eqref{eq:delta-perp} and \eqref{eq:delta-par}, far away
from the spike ($\left|v\right|\gg1$) we have\begin{eqnarray}
\delta_{\bot}\vec{X}\left(x\right) & = & e^{iKx-i\sqrt{K^{2}-1}t}\,\vec{m}\left[\gamma\left(cK-W\right)+i\tanh\left(\gamma\left(t-cx\right)\right)\right],\label{eq:example-S-mode}\\
\delta_{||}\vec{X}\left(x\right) & = & e^{iKx-i\sqrt{K^{2}-1}t}\,\vec{n}\left[\gamma\left(cK-W\right)+i\tanh\left(\gamma\left(t-cx\right)\right)\right],\nonumber \end{eqnarray}
and $\delta X^{1}=\delta X^{2}=0$ for both.%
\footnote{To obtain this, note that $K,W$ and $k,\omega$ are related by $K=-\gamma\left(ck+\sqrt{k^{2}-1}\right)$
and $W=-\gamma\left(k+c\sqrt{k^{2}-1}\right)$, from \eqref{eq:defn-KW}
and \eqref{eq:defn-uv}.%
} Fixing $t=0$ and evaluating at large distance $x=\pm\frac{L}{2}$,
they both become\[
\delta\vec{X}\left(\pm\frac{L}{2}\right)=e^{\pm iK\frac{L}{2}\pm i\delta_{\pm}}A_{\pm},\]
where the phase shifts and amplitudes at the two ends are given by
\begin{align}
\tan\left(\delta_{\pm}\right) & =\frac{-1\mp\gamma\sqrt{1-K^{2}}}{\gamma cK}\,,\label{eq:phases+amplitudes}\\
A_{\pm} & =\sqrt{\left(\gamma cK\right)^{2}+\left(\gamma\sqrt{1-K^{2}}\pm1\right)^{2}}\,.\nonumber \end{align}

The next step would be to impose periodic boundary conditions on $\delta X$
at $x=\pm\frac{L}{2}$. But here we encounter a problem, as the modes
have different amplitudes at the two ends.%
\footnote{Recall that the worldsheet velocity of the single spike is $1/c>1$.
Thus $(x,t)=(\pm L/2\,,0)$ might be better thought of as points before
and after the spike, rather than left and right of it. Consider instead
points $(x,t)$ with large $\left|t\right|$, for which both of the
modes $\delta_{\perp}$ and $\delta_{\vert\vert}$ in \eqref{eq:example-S-mode}
become\begin{align*}
\delta X & =e^{iKx-iWt}\left(\,\gamma\left(cK-W\right)+i\sign(t)\,\right)\\
 & =e^{iKx+\sqrt{1-K^{2}}t}\left(\,\gamma cK-i\gamma\sqrt{1-K^{2}}+i\sign(t)\,\right)\,.\end{align*}
In the second line we've chosen to focus on the growing mode $W=+i\sqrt{1-K^{2}}$.
Averaging over $x$ by taking the modulus, we get\[
\left|\delta X\right|=e^{\sqrt{1-K^{2}}t}\sqrt{\left(\gamma cK\right)^{2}+\left(\sign(t)-\gamma\sqrt{1-K^{2}}\right)^{2}}.\]
This is an exponentially growing mode, with a step in it where the
spike happens.%
} Instead we will demand only that the phases match at $x=\pm\frac{L}{2}$,
and allow the amplitudes to be different. (We will discuss this further
in the next section.) Then $K$ has to obey \[
KL+\delta_{+}\left(K\right)+\delta_{-}\left(K\right)=K_{n}L,\]
where $K_{n}=\frac{2\pi n}{L}$ is still the discretised momentum
of the vacuum solution. Taking $L$ very large we can approximate
$K$ by\[
K=K_{n}-\frac{1}{L}\delta\left(K_{n}\right)+\mathcal{O}(\frac{1}{L^{2}})\]
where $\,\delta(K)\equiv\delta_{+}(K)+\delta_{-}(K)$. Finally we
can determine the imaginary correction to the energy of the spike
from the four tachyonic modes, by putting $L\to\infty$:\begin{align}
\Delta E_{\mathrm{spike}} & =4\sum_{K}\frac{1}{2}W\left(K\right)\nonumber \\
 & \approx4\frac{L}{2\pi}\int_{-1}^{1}dK\,\frac{1}{2}W\left(K-\frac{1}{L}\delta(K)\right)\quad\mbox{as }L\to\infty\nonumber \\
 & =\Delta E_{\mathrm{hoop}}-i2\sqrt{\frac{1-c}{1+c}}.\label{eq:DeltaE-spike}\end{align}
In the expression above, $\Delta E_{\mathrm{hoop}}=iL/2$ is the correction
\prettyref{eq:DeltaE-vac} to the vacuum solution. Thus in the difference
$\Delta E_{\mathrm{spike}}-\Delta E_{\mathrm{hoop}}$ the IR divergence
from $L\to\infty$ is cancelled.

\subsection{About these boundary conditions\label{sub:About-these-boundary}}

We found that when $\left|K\right|<1$ the amplitude of the mode \prettyref{eq:example-S-mode}
is different at large positive and negative $x$. This is the obstruction
to imposing periodic boundary conditions, which we avoided by matching
only the phases. One should not be surprised that we cannot impose
periodic boundary conditions: they amount to gluing the string to
itself after some large number of windings, or rather, gluing the
vibrations on it to themselves, and this might not be allowed. 

For the giant magnon, one has to glue a series of magnons together
with $\sum_{i}p_{i}=0$ to obtain a valid closed string solution.
But is not clear that this is a condition on the allowed series of
single spikes. It would tell you about periodicity of the spatial
$X^{i}(x,t)$ under $t$, but say nothing about their behaviour at
large $\left|x\right|$. 

Here we consider a solution of two widely separated spikes with opposite
velocities $\frac{1}{c}$ and $-\frac{1}{c}$, because for this choice
we can impose honest boundary conditions. In this case we recover
the twice the energy correction \prettyref{eq:DeltaE-spike} obtained
above, one for each spike. This justifies our use of these unusual
boundary conditions.

\subsubsection{Two spikes}

As $x\to\pm\frac{L}{2}$, the amplitude of the mode \prettyref{eq:example-S-mode}
becomes $A_{\pm}$, given in \eqref{eq:phases+amplitudes}. This formula
is valid for $c>0$; for $c<0$ the sign $\pm$ is reversed, and we
have instead $\left|\delta X_{c<0}(\pm\frac{L}{2},0)\right|=A_{\mp}$. 

This immediately suggests the following way to impose consistent boundary
conditions: take two spikes, far apart, with parameters $c$ and $-c$.
Each is in a box of length $L$, and we connect these together. That
is, consider \[
X^{\mu}(x,t)=\begin{cases}
X_{\mathrm{spike}(c)}\left(x-\frac{L}{2}\,,t\right) & \mbox{for }\quad0<x<L\,,\\
X_{\mathrm{spike}(-c)}\left(x-\frac{3L}{2}\,,t\right) & \qquad\: L<x<2L\end{cases}\]
which is an approximate solution near $t=0$. In fact it is a part
of a scattering solution, since the two spikes have velocities $1/c$
and $-1/c$. It can be viewed as an excitation above a hoop of length
$2L$.

Vibrations of this solution will be described by the same modes we
have been using, and we again focus on the $\left|K\right|<1$ sphere
modes, which give the imaginary energy correction. For the boundary
condition at $x=L$ , both modes $\delta X$ have amplitude $A_{+}$,
so matching them sets their phases equal there. And at $x=0,2L$ we
can impose periodic boundary conditions, since both modes have amplitude
$A_{-}$ there. The resulting condition on the allowed $K$ is simply
\[
K=K_{n}-\frac{1}{2L}\delta_{(c)}\left(K_{n}\right)-\frac{1}{2L}\delta_{(-c)}\left(K_{n}\right)+\mathcal{O}(\frac{1}{L^{2}})\,,\]
where $K_{n}=\frac{2\pi n}{2L}$ are now the allowed wave numbers
for the vacuum in length $2L$. This leads to energy correction\begin{align*}
\Delta E & =\Delta E_{\mathrm{spike}(c)}+\Delta E_{\mathrm{spike}(-c)}\,,\end{align*}
i.e. we obtain the sum of the corrections we calculated in \eqref{eq:DeltaE-spike}
by imposing our phase-only boundary condition at $x=\pm\frac{L}{2}$.
The finite piece (after subtracting the vacuum's $\Delta E_{\mathrm{hoop}}$)
is twice the finite piece for one spike.

\section{Conclusion\label{sec:Conclusion}}

In this paper we determined the bosonic and fermionic modes of the
single spike solution. Because there is a mismatch between the modes
in these two sectors, both in number and in their masses, the spike
cannot be supersymmetric. Some of the bosonic modes are tachyonic,
showing that the single spike is unstable, like the relevant `vacuum'
solution which we referred to as the hoop.

We found that the Hamiltonian for small fluctuations of this vacuum
is $\Delta-\Phi$. The winding $\Phi$ has replaced the angular momentum
$J$ found in the Hamiltonian for the magnon case, which is not surprising
given that T-duality relates similar solutions in flat space. Using
this result we performed a semi-classical calculation of the lifetime
of the solution.

The dispersion relation for giant magnons \eqref{eq:charges-for-magnon}
is periodic in $p$, which is the signature of discrete space. This
is understood to be the position along a spin chain. One should not
read the apparent lack of such periodicity in the single spike's case
\eqref{eq:charges-for-spike} as evidence against such discreteness.
The recent paper \cite{Ishizeki:2007kh} allows $p$ outside our range
$0<p<\pi$, and finds that $\Delta-\Phi$ becomes periodic (their
figure 1). However it is not clear that for the single spike this
parameter $p$ can still be interpreted as a spin-chain momentum. 

It had been conjectured that the single spike is dual to an excitation
of an anti-ferromagnetic spin chain. \cite{Ishizeki:2007we,Roiban:2006jt}
These have been various attempts to find an $N$-body description
of the giant magnon, such as a Hubbard models, \cite{Rej:2005qt,Drummond:2007gt}
as was done for sine-gordon kinks. \cite{Babelon:1993bx,Ruijsenaars:1986vq}
It is possible that this solution will be another test case for such
a description.

The single spike is an excitation of an unstable vacuum state, the
string wrapped around an equator of $S^{5}$. One can stabilise such
loops of string by making them rotate in other planes. \cite{Frolov:2003qc,Frolov:2003xy}
These can carry large angular momentum by being wound many times.
It is possible that adding these extra angular momenta may stabilize
the spike solution too, and it may be this object which has a more
natural gauge theory dual.

\section*{Acknowledgements}

We would like to thank Marcus Spradlin, Antal Jevicki and Georgios
Papathanasiou for numerous conversations. Robert de Mello Koch, Alin
Tirziu and Martin Kruczenski were kind enough to read drafts of the
paper for us. Michael Abbott would also like to thank Jeff Murugan
for making him read \cite{Hofman:2006xt}. 

Our research was supported in part by DOE grant DE-FG02-91ER40688-Task
A. In\^{e}s~Aniceto was also supported by POCI 2010 and FSE, Portugal,
through the fellowship SFRH/BD/14351/2003.

\bigskip \bigskip \bigskip 

\appendix

\section{Fermionic Zero modes\label{sec:Fermionic-Zero-modes}}

This appendix contains the exact analogue of the calculation done
for the magnon in \cite{Minahan:2007gf}. It has the virtue that it
is easy to see why the single spike has twice as many modes as the
magnon does; this was somewhat obscure in the non-zero mode calculations
of section \ref{sub:fermi-nonzero-modes}. But the result is identical
to simply setting $\omega=0$ there.

The zero modes are those with $\partial_{u}\Psi^{I}=0$. Then the
second-order equation \eqref{eq:second-order Theta eq} becomes \begin{align*}
\left(\frac{1}{\gamma\cos\theta}\mathcal{D}_{v}\frac{1}{\gamma\cos\theta}\mathcal{D}_{v}+1\right)\Psi^{1} & =0,\end{align*}
which factorises, and that is why the calculations are much easier
than the non-zero modes. 

This equation implies that $\left(\mathcal{D}_{v}-\eta\, i\gamma\cos\theta\right)\Psi^{1}=0$
with one of $\eta=\pm1$, or pulling $(\rho_{0}-\rho_{1})$ out: \[
(\rho_{0}-\rho_{1})\left\{ \partial_{v}+\frac{1}{2}G\Gamma_{\phi\theta}+\eta\: i\gamma\cos\theta\right\} \Theta^{1}=0.\]
As for the non-zero modes, we first ignore the $\kappa$-symmetry
projection and solve for $\Theta^{1}$ alone. The matrix part of this
equation involves only $\mathbf{1}$ and $\Gamma_{\phi\theta}$, which
can be simultaneously diagonalised. Write the solution as \begin{align*}
\Theta^{1} & =\Theta_{+}+\Theta_{-}=f_{+}(v)U_{+}+f_{-}(v)U_{-}\,,\end{align*}
where the spinors $U_{\pm}$ (and so $\Theta_{\pm}$) are $\Gamma_{\phi\theta}$
eigenvectors, with eigenvalues $\pm i$ respectively. All that is
left to solve is \[
\left\{ \partial_{v}\pm\frac{i}{2}G+\eta\, i\gamma\cos\theta\right\} f_{\pm}(v)=0.\]
The solutions are pure phase, \begin{align*}
f_{\pm}(v) & =e^{\pm i\chi}e^{i\eta\chi_{2}} & \mbox{where }\quad e^{i\chi} & =\left(\frac{\sinh v+ic}{\sinh v-ic}\right)^{1/4}\sqrt{\tanh v+i\sech v},\\
 &  & e^{\pm i\chi_{2}} & =\sech v\pm i\tanh v.\end{align*}
The difference between these solutions and the giant magnon's ones
\cite{Minahan:2007gf} is that instead of a modulating factor $\sech u$,
we get an extra phase $e^{i\eta\chi_{2}}$. It is this modulating
factor which makes one solution normalisable, and allows Minahan to
reject the other sign of $\eta$ for producing a solution which diverges
at large $u$. But in our case both signs lead to non-normalisable
solutions. The general solution is a linear combination of the $\eta=+1$
and $\eta=-1$ cases: \begin{align*}
\Psi^{1} & =-i(\rho_{0}-\rho_{1})\frac{1}{\sqrt{1-c}}\sum_{\pm}e^{\pm i\chi}\sum_{\eta}e^{i\eta\chi_{2}}U_{\pm}^{\eta}\,.\end{align*}
(We've smuggled in a factor of $\sqrt{1-c}$ for reasons of aesthetic
balance between between $\Psi^{1}$ and $\Psi^{2}$.) Writing out
$(\rho_{0}-\rho_{1})$ and using the identity $e^{\pm2i\chi}=(p-r)\mp i(q-s)$,
we obtain:\begin{align*}
\Psi^{1} & =\frac{i}{\sqrt{1-c}}\left[\Gamma_{0}\left(\cos\chi+\Gamma_{\phi\theta}\sin\chi\right)-\Gamma_{\phi}\left(\cos\chi-\Gamma_{\phi\theta}\sin\chi\right)\right]\left(\sech v\, U_{0}+\tanh v\,\tilde{U}_{0}\right)\end{align*}
where we've combined the arbitrary spinors $U_{\pm}^{\eta}$ into\begin{align*}
U_{0} & =-\left(U_{+}^{\eta=1}+U_{-}^{\eta=1}\right)-\left(U_{+}^{\eta=-1}+U_{-}^{\eta=-1}\right),\\
\tilde{U}_{0} & =-i\left(U_{+}^{\eta=1}+U_{-}^{\eta=1}\right)+i\left(U_{+}^{\eta=-1}+U_{-}^{\eta=-1}\right).\end{align*}
The reason for this choice is that the Majorana condition $\Psi^{*}=\Psi$
now simply requires that $U_{0}$ and $\tilde{U}_{0}$ be themselves
Majorana spinors. (The $\Gamma$-matrices are all imaginary, thus
$\Gamma_{\phi\theta}$ is real.)

Having found $\Psi^{1}$, we immediately have $\Psi^{2}$ as an operator
acting on it, from \eqref{Psi2 in terms of Psi1}, with no further
choices to make. We can write:\begin{align*}
\Psi^{2} & =\frac{\Gamma_{\star}\Gamma_{\theta}}{\sqrt{1+c}}\left[\Gamma_{0}\left(\cos\tilde{\chi}+\Gamma_{\phi\theta}\sin\tilde{\chi}\right)-\Gamma_{\phi}\left(\cos\tilde{\chi}-\Gamma_{\phi\theta}\sin\tilde{\chi}\right)\right]\left(\sech v\,\tilde{U}_{0}-\tanh v\, U_{0}\right)\end{align*}
where as before $e^{i\tilde{\chi}}=e^{-i\chi+i\chi_{2}}$, and $(r\pm i\, s)=\pm i\gamma\cos\theta e^{\pm i(\tilde{\chi}-\chi)}$
was used.

Comparing these zero modes with the non-zero modes \eqref{eq:nonzero psi1}
and \eqref{eq:nonzero psi2}, it is clear that they are simply the
$\omega=0$ case of the latter (i.e. $\alpha=\beta=0$). This is different
from the supersymmetric giant magnon case, where the massive non-zero
modes of \cite{Papathanasiou:2007gd} do not connect to the zero modes
of \cite{Minahan:2007gf}.

Let us pause to count these modes: the four spinors $U_{\pm}^{\eta}$
are $\Gamma_{\phi\theta}$-eigenspinors, thus have 16 complex components
each. They must be Weyl spinors, i.e. $\Gamma_{11}$-eigenspinors,
which cuts the number in half. Requiring $U_{0}$ and $\tilde{U}_{0}$
to be Majorana cuts it in half again, to 16 components in total. This
is the same number we would count by setting $\omega=0$ in the non-zero
modes above, as must be the case: $U_{0}$ is the same spinor. At
this stage \cite{Minahan:2007gf} had 8 complex components. The argument
below cuts it by another factor of 2 in both cases.

\subsubsection{Slow-motion }

In \cite{Minahan:2007gf}, Minahan uses an argument which runs as
follows: regard the spinors $U_{0}$ and $\tilde{U}_{0}$ as a moduli
of the solution, and allow them to become time-dependent, $\partial_{u}U\neq0$.
Substituting a zero mode into the action will give zero, but this
`slowly-moving' mode needn't do so. The zero modes whose slowly-moving
cousins give a non-zero action are `real' zero modes, the others pure
gauge. \cite{Manton:1981mp}

When substituting the slowly-moving mode $\Theta=\sum F(v)U(u)$ into
the Lagrangian, the equations of motion force everything except the
$\partial_{u}$ terms to vanish, leaving\[
\mathcal{L}_{F}=-i\gamma(1-c)\:\overline{\Theta}^{1}(\rho_{0}-\rho_{1})\partial_{u}\Theta^{1}+i\gamma(1+c)\:\overline{\Theta}^{2}(\rho_{0}+\rho_{1})\partial_{u}\Theta^{2}.\]
(As usual $\overline{\Theta}=\Theta^{\dagger}\Gamma_{0}$.) Using
the identities $2\Gamma_{0}(\rho_{0}\pm\rho_{1})=-(\rho_{0}\pm\rho_{1})^{\dagger}(\rho_{0}\pm\rho_{1})$
this becomes\[
\mathcal{L}_{F}=i\gamma\frac{1-c}{2}\Psi^{1\dagger}\partial_{u}\Psi^{1}-i\gamma\frac{1+c}{2}\Psi^{2\dagger}\partial_{u}\Psi^{2}.\]

Now plug in $\Psi^{I}$ from above, to obtain\begin{align*}
\mathcal{L}_{F} & =\frac{i\gamma}{2}\left[(\Gamma_{0}-\Gamma_{\phi})\left(\sech v\, U_{0}+\tanh v\,\tilde{U}_{0}\right)\right]^{\dagger}\partial_{u}\left[(\Gamma_{0}-\Gamma_{\phi})\left(\sech v\, U_{0}+\tanh v\,\tilde{U}_{0}\right)\right]\\
 & \quad-\frac{i\gamma}{2}\left[(\Gamma_{0}-\Gamma_{\phi})\left(\sech v\,\tilde{U}_{0}-\tanh v\, U_{0}\right)\right]^{\dagger}\partial_{u}\left[(\Gamma_{0}-\Gamma_{\phi})\left(\sech v\,\tilde{U}_{0}-\tanh v\, U_{0}\right)\right].\end{align*}
Both $U_{0}$ and $\tilde{U}_{0}$ always appear acted on by $(\Gamma_{0}-\Gamma_{\phi})$.
Thus only those modes satisfying $(\Gamma_{0}+\Gamma_{\phi})U_{0}=0$
and $(\Gamma_{0}+\Gamma_{\phi})\tilde{U}_{0}=0$ contribute. The situation
is identical to the magnon case in that only half of the modes appearing
in $\Psi^{I}$ contribute here (`are real,' meaning true, zero modes).
But since there are two constant Majorana--Weyl spinors $U_{0}$ and
$\tilde{U}_{0}$ here, instead of only one for the giant magnon, there
are twice as many modes: 8 instead of 4 complex degrees of freedom.
Thus for both the non-zero modes and the zero modes, in the fermionic
sector, we find twice as many as in the giant magnon case.

\section{Hamiltonian Formulation and Energy Corrections\label{sec:Appendix-Energy-Corr}}

\subsection{Quadratic 2-dimensional Hamiltonian\label{sub:2-dimensional-Hamiltonian}}

Starting from the Lagrangian for the fluctuations (\ref{eq:Lagr fluctuations}),
we find its quadratic part to be (up to factors of $\lambda$):\[
\tilde{\mathcal{L}}^{2}=\frac{1}{2}\left(-\partial^{a}\tilde{\tau}\partial_{a}\tilde{\tau}+\partial^{a}\tilde{\eta}_{k}\partial_{a}\tilde{\eta}_{k}+\partial^{a}\tilde{\phi}\partial_{a}\tilde{\phi}+\partial^{a}\tilde{\theta}_{s}\partial_{a}\tilde{\theta}_{s}+\tilde{\eta}_{k}\tilde{\eta}_{k}-\tilde{\theta}_{s}\tilde{\theta}_{s}\right).\]
By determining the conjugate momenta for each of the fluctuation fields,
$\tilde{\Pi}_{\mu}=\frac{\partial\tilde{\mathcal{L}}}{\partial\left(\partial_{0}\tilde{X}^{\mu}\right)}$,
we find\begin{align*}
\tilde{\Pi}_{\tilde{\tau}} & =\partial_{0}\tilde{\tau}\\
\tilde{\Pi}_{\tilde{X}^{\mu}} & =-\partial_{0}\tilde{X}^{\mu}\quad\mbox{for }\tilde{X}^{\mu}=\tilde{\eta}_{k},\tilde{\theta}_{s},\tilde{\phi}.\end{align*}
From these we can construct the corresponding Hamiltonian density
in the usual way, obtaining\[
\tilde{\mathcal{H}}^{2}=\frac{1}{2}\left(-\tilde{\Pi}_{\tilde{\tau}}^{2}+\tilde{\Pi}_{\tilde{\phi}}^{2}+\tilde{\Pi}_{\tilde{\eta}_{k}}^{2}+\tilde{\Pi}_{\tilde{\theta}_{s}}^{2}+\tilde{\eta}_{k}\tilde{\eta}_{k}-\tilde{\theta}_{s}\tilde{\theta}_{s}\right).\]

We want to check that the quantity $\Delta-\Phi$ is just this Hamiltonian.
To do so we start by determining the Hamiltonian corresponding to
the original bosonic Lagrangian (\ref{eq:string lagrang}), and expand
it in fluctuations. The conjugate momenta for the fields are given
by $\Pi_{\mu}=\frac{\partial\mathcal{L}}{\partial\left(\partial_{0}X^{\mu}\right)}$
where $X^{\mu}=\tau,\phi,\eta_{k},\theta_{s}$. To find the Hamiltonian
for the fluctuations, we expand the fields as in (\ref{eq:expansion in fluct}),
as well as the momenta: \begin{eqnarray*}
\Pi_{\mu} & = & \Pi_{\mu}^{cl}+\lambda^{-\frac{1}{4}}\tilde{\Pi}_{\mu}\quad;\qquad X^{\mu}=X_{cl}^{\mu}+\lambda^{-\frac{1}{4}}\tilde{X}^{\mu},\end{eqnarray*}
where the classical values of the fields are $\Pi_{\tau}^{cl}=1$,
$\Pi_{X^{\mu}\ne\tau}^{cl}=0$, $\tau_{cl}=t,\,\phi_{cl}=x$ and all
other fields are zero. The expansion of the Hamiltonian then gives:\begin{eqnarray*}
\mathcal{H}_{b} & = & \frac{1}{2\sqrt{\lambda}}\left(-\tilde{\Pi}_{\tilde{\tau}}^{2}+\tilde{\Pi}_{\tilde{\phi}}^{2}+\tilde{\Pi}_{\tilde{\eta}_{k}}^{2}+\tilde{\Pi}_{\tilde{\theta}_{s}}^{2}-\left(\partial_{1}\tilde{\tau}\right)^{2}+\left(\partial_{1}\tilde{\phi}\right)^{2}+\left(\partial_{1}\tilde{\eta_{k}}\right)^{2}+\left(\partial_{1}\tilde{\theta_{s}}\right)^{2}\right)+\\
 &  & \qquad\qquad+\frac{1}{2\sqrt{\lambda}}\left(\tilde{\eta}_{k}\tilde{\eta}_{k}-\tilde{\theta}_{s}\tilde{\theta}_{s}\right)-\frac{1}{\lambda^{\frac{1}{4}}}\left(\tilde{\Pi}_{\tilde{\tau}}-\left(\partial_{1}\tilde{\phi}\right)\right)+\mathcal{O}\left(\frac{1}{\lambda}\right).\end{eqnarray*}
The Virasoro constraint (\ref{eq:Virasoro const}) is equivalent to
setting $\mathcal{H}_{b}=0$.

It is easy to check that $\Delta-\Phi$ can be written in terms of
the fields and conjugate momenta as\[
\Delta-\Phi=\frac{\sqrt{\lambda}}{2\pi}\int dx\left(\frac{1}{\lambda^{\frac{1}{4}}}\left(\tilde{\Pi}_{\tilde{\tau}}-\left(\partial_{1}\tilde{\phi}\right)\right)\right).\]
By using the Virasoro constraint in the form $\mathcal{H}_{b}=0$,
we finally find \begin{eqnarray*}
\Delta-\Phi & = & \int\frac{dx}{2\pi}\Big(-\tilde{\Pi}_{\tilde{\tau}}^{2}+\tilde{\Pi}_{\tilde{\phi}}^{2}+\tilde{\Pi}_{\tilde{\eta}_{k}}^{2}+\tilde{\Pi}_{\tilde{\theta}_{s}}^{2}-\left(\partial_{1}\tilde{\tau}\right)^{2}+\left(\partial_{1}\tilde{\phi}\right)^{2}+\left(\partial_{1}\tilde{\eta_{k}}\right)^{2}+\left(\partial_{1}\tilde{\theta_{s}}\right)^{2}\\
 &  & \qquad\qquad\qquad\qquad\qquad\qquad+\tilde{\eta}_{k}\tilde{\eta}_{k}-\tilde{\theta}_{s}\tilde{\theta}_{s}\Big),\end{eqnarray*}
which returns the expected expression, when we drop the gauge fluctuations.

\subsection{Modes for the hoop (vacuum) solution\label{sub:hoop-modes}}

It is simple to solve the equations of motion from the bosonic Lagrangian
$\mathcal{L}_{B}$ \eqref{eq:Lagr fluctuations} in order to determine
the modes for the hoop solution. The transverse modes are \begin{align*}
\tilde{\eta}_{k}(x,t) & =e^{iKx-iWt}f_{k}(K),\qquad W^{2}=K^{2}+1\,,\\
\tilde{\theta}_{s}(x,t) & =e^{iKx-iWt}g_{s}(K),\qquad W^{2}=K^{2}-1\,,\end{align*}
i.e $m^{2}=1$ in the $AdS$ directions, and $m^{2}=-1$ on the sphere,
the same masses as for the single spike's modes. The longitudinal
modes are massless:\begin{align*}
\tilde{\tau}(x,t) & =e^{iKx-i\left|K\right|t}f(K),\\
\tilde{\phi}(x,t) & =e^{iKx-i\left|K\right|t}g(K).\end{align*}

The same modes can also be obtained from those for the single spike,
by going far away from the spike itself. The sphere modes $\delta_{\perp}$
\prettyref{eq:delta-perp} and $\delta_{\vert\vert}$ \prettyref{eq:delta-par}
of section \ref{sub:bosonic-nonzero-modes} become these simple ones
$\tilde{\theta}_{s}$ in the limit $v\to\infty$, and the $AdS$ modes
are identical. The $\tilde{\phi}$ mode is the $v\to\infty$ limit
of $\delta_{r}$ \eqref{eq:delta-r}, now more obviously pure gauge.
We did not write down the analogue of the $\tilde{\tau}$ mode (among
the spike's non-zero modes) as we were focusing on the spatial part,
but this too is pure gauge. 

Performing the same limit $v\to\infty$ for the fermionic modes \eqref{eq:nonzero psi1}
and \eqref{eq:nonzero psi2} leaves the following modes for the hoop:\begin{align*}
\Psi^{1} & =\frac{i}{\sqrt{1-c}}\left[\Gamma_{0}\left(\cos\chi+\Gamma_{\phi\theta}\sin\chi\right)-\Gamma_{\phi}\left(\cos\chi-\Gamma_{\phi\theta}\sin\chi\right)\right]\\
 & \qquad\qquad\times\left(\cos\beta\,\tilde{U}_{0}+\sin\beta\,\Gamma_{\phi\theta}\tilde{U}_{1}\right),\\
\Psi^{2} & =\frac{-1}{\sqrt{1+c}}\Gamma_{*}\Gamma_{\theta}\left[\Gamma_{0}\left(\cos\tilde{\chi}+\Gamma_{\phi\theta}\sin\tilde{\chi}\right)-\Gamma_{\phi}\left(\cos\tilde{\chi}-\Gamma_{\phi\theta}\sin\tilde{\chi}\right)\right]\\
 & \qquad\qquad\times\frac{1}{1+4\omega^{2}}\left(\cos\tilde{\beta}\, U_{0}+\sin\tilde{\beta}\,\Gamma_{\phi\theta}U_{1}\right).\end{align*}

\bibliographystyle{utphys}
\bibliography{spiky}

\end{document}